\documentclass[apj]{emulateapj}

\usepackage{graphicx}
\usepackage{amssymb}
\usepackage{amsmath}
\usepackage{multirow}
\usepackage{float}
\usepackage{comment}

\usepackage{natbib}

\newcommand{\be}{\begin{equation}}
\newcommand{\ee}{\end{equation}}

\begin{document}
 
\title{Geminga's puzzling pulsar wind nebula}
\author{B. Posselt}
\affil{Department of Astronomy \& Astrophysics, Pennsylvania State University, 525 Davey Lab,University Park, PA 16802, USA }
\email{posselt@psu.edu}
\author{G. G. Pavlov}
\affil{Department of Astronomy \& Astrophysics, Pennsylvania State University, 525 Davey Lab,University Park, PA 16802, USA}

\author{P. O. Slane}
\affil{Harvard-Smithsonian Center for Astrophysics, 60 Garden Street, Cambridge, MA 02138, USA}

\author{R. Romani}
\affil{Department of Physics, Stanford University, Stanford, CA 94305, USA}

\author{N. Bucciantini}
\affil{INAF Osservatorio Astrofisico di Arcetri, L.go E. Fermi 5, I-50125 Firenze, Italy; INFN Sezione di Firenze, Via G. Sansone 1, I-50019 Sesto (Firenze), Italy}

\author{A. M. Bykov}
\affil{A.F. Ioffe Physical-Technical Institute, St. Petersburg 194021, St.Petersburg State Politechnical University, Russia}
\affil{International Space Science Institute, Bern, Switzerland}

\author{O. Kargaltsev}
\affil{Department of Physics, The George Washington University, 725 21st St, NW, Washington, DC 20052, USA}

\author{M. C. Weisskopf}
\affil{NASA/Marshall Space Flight Center, ZP12, 320 Sparkman Drive, Huntsville, AL 35805, USA}

\author{C.-Y. Ng}
\affil{Department of Physics, The University of Hong Kong, Pokfulam Road, Hong Kong, China}

\begin{abstract}
We report on six new {\sl{Chandra}} observations of the Geminga pulsar wind nebula (PWN). The PWN consists of three distinct elongated structures -- two $\approx 0.2 d_{250}$\,pc long lateral tails and a segmented axial tail of $\approx 0.05 d_{250}$\,pc length, where $d_{250}=d/(250 {\rm pc})$. 
The photon indices of the power law spectra of the lateral tails, $\Gamma \approx 1$, are significantly harder than those of the pulsar ($\Gamma \approx 1.5$) and the axial tail ($\Gamma \approx 1.6$). 
There is no significant diffuse X-ray emission between the lateral tails -- the ratio of the X-ray surface brightness between the south tail and this sky area is at least 12. The lateral tails apparently connect directly to the pulsar and show indication of moving footpoints.
The axial tail comprises time-variable emission blobs. However, there is no evidence for constant or decelerated outward motion of these blobs.
Different physical models are consistent with the observed morphology and spectra of the Geminga PWN.  
In one scenario, the lateral tails could represent an azimuthally asymmetric shell whose hard emission is caused by the Fermi acceleration mechanism of colliding winds.
In another scenario, the lateral tails could be luminous, bent polar outflows, while the blobs in the axial tail could represent a crushed torus.
In a resemblance to planetary magnetotails, the blobs of the axial tail might also represent short-lived plasmoids which are formed by magnetic field reconnection in the relativistic plasma of the pulsar wind tail.
\end{abstract}


\section{Introduction}
\label{intro}
\begin{figure*}[th]
\noindent\begin{minipage}[b]{.45\textwidth}
\begin{center}
\includegraphics[width=85mm]{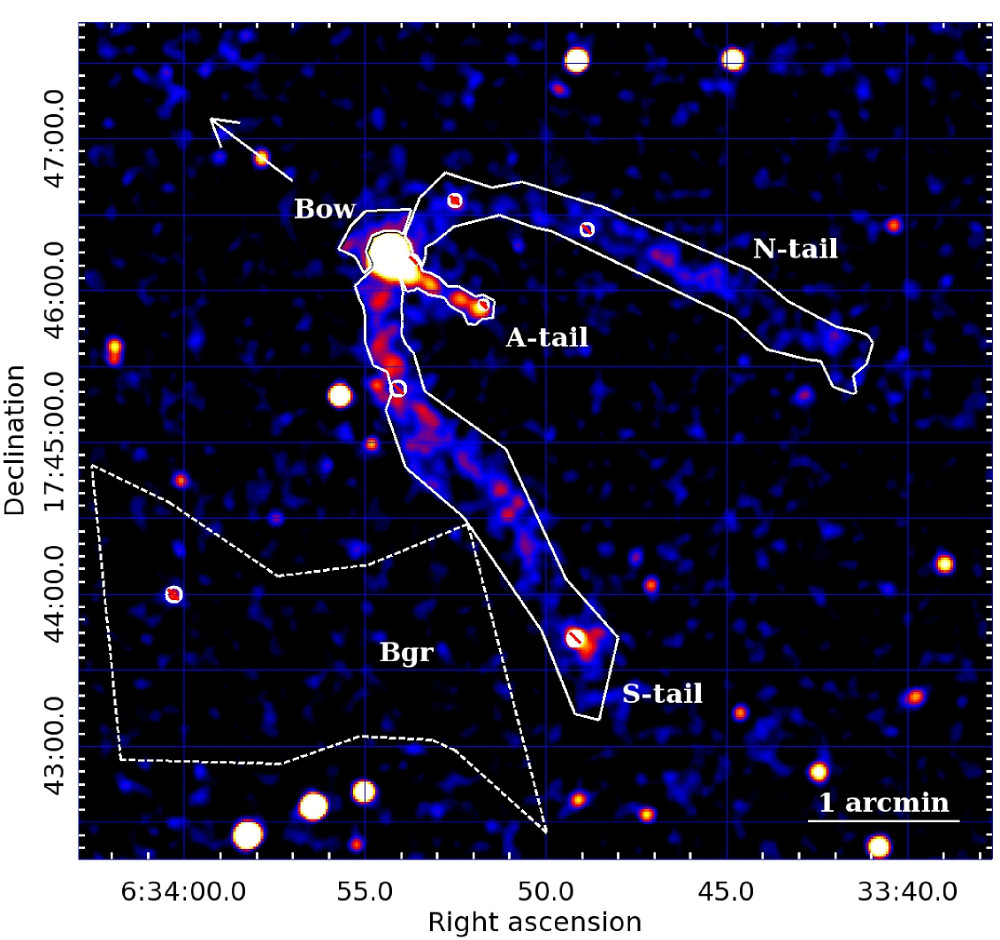}
\end{center}
\end{minipage} 
\begin{minipage}[b]{.55\textwidth}
\begin{center}
\includegraphics[width=95mm]{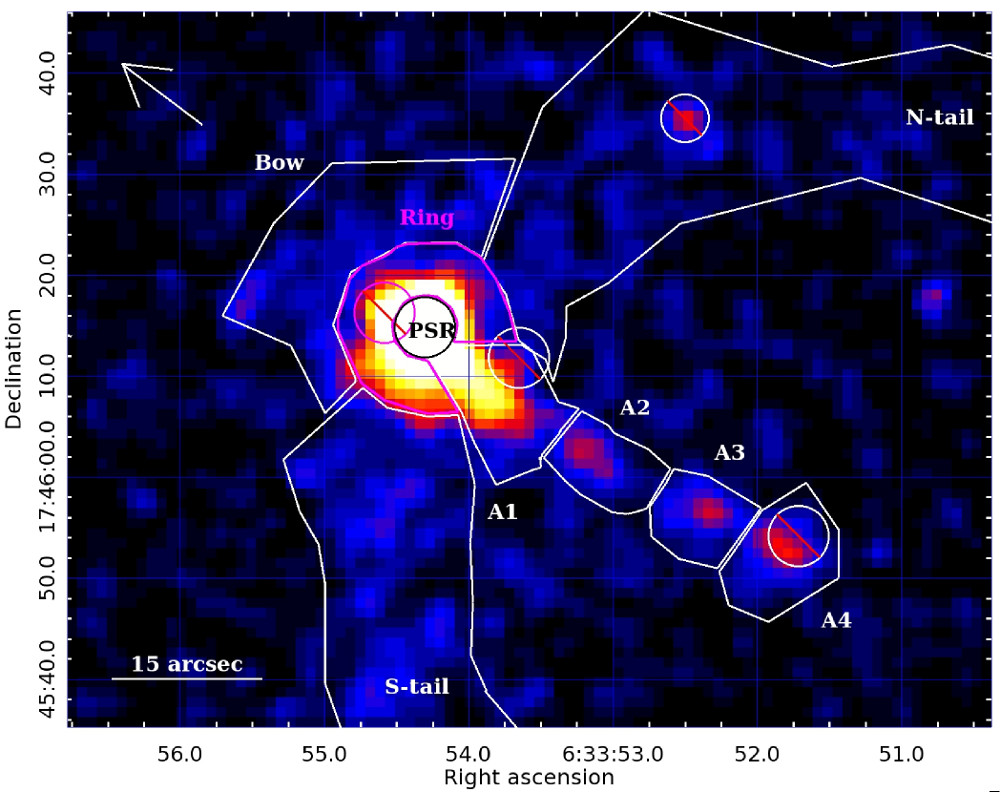}
\end{center}
\end{minipage}
\caption{The extended emission around Geminga and our nomenclature for the spectral extraction regions. The merged {\sl{Chandra}} exposure-map-corrected 0.3-7\,keV ACIS-I image was produced from 11 event files (2012-2013) as described in Section~\ref{dataana}. The image was binned to pixel sizes of $0\farcs{984}$ (2 ACIS pixels). The images were smoothed with a Gaussian with $\sigma_x =\sigma_y = 2$\,image pixels (\emph{left panel}) and 1\,image pixel (\emph{right panel}). The color scale was manually adapted to highlight the regions of interest. 
Crossed circles indicate excluded point sources that were identified in multi-wavelength data. The dashed region is the used background region. The arrow indicates the direction of the proper motion.}
\label{Overview}
\end{figure*}

The well-known $\gamma$-ray pulsar Geminga (PSR\,J0633+1746) has a rather unusual pulsar wind nebula (PWN). Previous XMM-$Newton$ and {\sl Chandra} observations revealed three tails -- two $\approx 2\arcmin$ long bent lateral (outer) tails and an $\approx 45\arcsec$ long axial tail with distinguishable variable blobs -- in addition to some emission in front of the pulsar (\citealt{Caraveo2003, Pavlov2006}, and  \citealt{Pavlov2010}, hereafter PBZ10); see Figure~\ref{Overview} for an updated image of the PWN.
Many of the $\sim 90$ PWNe detected in X-rays show symmetric torus-jet or bow-shock morphologies \citep{Kargaltsev2008}.
The radio to GeV emission of PWNe is synchrotron radiation which is produced by shocked relativistic pulsar wind and the ambient medium; for reviews see, e.g., \citet{Kargaltsev2012, 2011Slane, Gaensler2006}. The recent numerical magnetohydrodynamic simulations of {\rm PWN}e reproduce observed features related to {\rm PWN}e around young pulsars, e.g., torii and jets;  see, e.g., \citet{Porth2014,Bucciantini2011} and references therein. 
The `three-tail' PWN of Geminga is unusual. There is only one other known PWN partly resembling Geminga -- around the more energetic, but more distant PSR\,J1509$-$5850 (\citealt{Klingler2016}).\\

The radio-quiet Geminga is a middle-aged (0.1-1\,Myr) pulsar with a spin-down age $\tau_{SD}=340$\,kyr, a period $P=0.237$\,s, and a spin-down power $\dot{E}=3.3 \times 10^{34}$\,erg\,s$^{-1}$ \citep{Bertsch1992}. \textit{Fermi} has discovered many  radio-quiet and radio-loud $\gamma$-ray pulsars with similar spin-down properties (e.g., \citealt{Abdo2013}). For example, there are 17 similar middle-aged $\gamma$-ray pulsars within 2\,kpc. Geminga, however, is one of the closest -- with a distance of only $d=250^{+230}_{-80}$\,pc (corrected for the Lutz-Kelker bias; \citealt{Verbiest2012,Faherty2007}). 
Though the nominal distance error of Geminga encompasses a large distance range, there are several arguments for a distance smaller than 700\,pc, more likely even smaller than 500\,pc: neutron star radius constraints and a low value for the column density of the absorbing interstellar medium ($d<500$\,pc; \citealt{Mori2014,Kargaltsev2005,Lallement2014}), the pulsar's $\gamma$-ray efficiency ($d<300$\,pc or $d<700$\,pc depending on emission model; \citealt{Abdo2010Geminga}), and a low dispersion measure in the recently claimed low-frequency radio detection (220\,pc  \citep{Malov2015} for the NE2001 model \citealt{Cordes2002}).\\

\citet{Faherty2007} measured a proper motion of $178.2 \pm 1.8$\,mas\,yr$^{-1}$ for Geminga. This corresponds to a transverse velocity $v_t \approx 211 d_{250}$\,km\,s$^{-1}$, where $d_{250}=d/(250 {\rm pc})$. Since this velocity is a factor of about 10 larger than the typical sound speed in the interstellar medium, one would expect a bowshock-tail PWN around Geminga (PBZ10). Geminga's proximity provides the opportunity to probe PWN models in detail.\\

Based on the results of the previous X-ray observations of Geminga's PWN, PBZ10 proposed two possible physical scenarios:
The outer tails could be a sky projection of a limb-brightened shell formed in the region of the contact discontinuity (CD) which separates the shocked PWN from the shocked ambient medium. 
The axial tail could then be interpreted as a jet launched along the pulsar's spin axis.
In the second scenario, the lateral tails could be polar outflows from the pulsar bent by the ram pressure from the ambient interstellar medium (ISM). In this case, the axial tail could be a shocked equatorial outflow collimated by ram pressure.
These two scenarios imply different geometries. In the first scenario, the axial tail marks roughly the (projected) direction of the rotation axis, while in the second scenario the projected rotation axis would be marked by the footpoints of the lateral tails.\\  

Constraints on the geometry of Geminga can be obtained from its $\gamma$-ray and X-ray pulse shape.
\citet{Pierbattista2014} used four different magnetosphere models (PC -- Polar Cap, SG -- Slot Gap, OG -- Outer Gap, OPC -- One Pole Caustic) to fit radio  and $\gamma$-ray pulse profiles of many $\gamma$-ray pulsars. Generally, all the considered models perform poorly, in particular if both radio and $\gamma$-rays are considered together.
For the radio-quiet Geminga pulsar, the best fitting (SG) model used by \citet{Pierbattista2014} resulted in $\alpha=42^\circ$, $\zeta=51^\circ$, where the obliquity $\alpha$ is the angle between the rotational and magnetic axis, and $\zeta$ is the angle between the direction of sight and the rotation axis. Even this best-fitting model had problems with Geminga's $\gamma$-ray interpulse which it poorly reproduced.
Previously, \citet{Watters2009} obtained $\alpha=10^\circ-25^\circ$, $\zeta=85^\circ$ for the OG-model, and two solutions for the two-pole caustic magnetosphere model ($\alpha=30^\circ-80^\circ$, $\zeta=90^\circ$, and $\alpha=90^\circ$, $\zeta=55^\circ-80^\circ$).
\citet{Malov1998}, on the other hand, argued that the radio-quietness of Geminga indicates an aligned rotator, i.e., small $\alpha$.
The currently discussed models of Geminga's pulse shape seem to agree that  $\zeta>50^\circ$ -- with an uncertain value of $\alpha$. \\

The proposed two physical scenarios for Geminga's PWN lead to different expectations for temporal changes whose reality can be checked. If the blobs of the axial tail are inhomogeneities in a jet, one would expect outward motion and potentially softening further away from the pulsar. 
In addition, in the case of a shell explanation for the lateral tails, the space between these limb-brightened shell boundaries should be filled with faint X-ray emission which should become detectable in a sufficiently long observation.
In this paper, we report on the analysis of six epochs of recently acquired deep Geminga observations. We investigate the merged data set as well as the observation at individual epochs to probe the different physical interpreations of Geminga's unusual PWN and also to constrain Geminga's obliquity and orientation with respect to proper motion and line of sight.   

\section{Observations and Data Analysis}
\label{dataana}
We observed Geminga with {\sl{Chandra}} ACIS-I in six individual epochs, each comprising about 100\,ks exposure time, from November 2012 to September 2013. The `very faint' telemetry mode was used. 
A list of our new and the previous {\sl{Chandra}} observations is given in Table~\ref{obslog}.\\
\begin{deluxetable}{lrcccr}[bh]
\tablecaption{Observations\label{obslog}}
\tablewidth{0pt}
\tablehead{
\colhead{Epoch} & \colhead{ObsID} &\colhead{Date} & \colhead{MJD} & \colhead{Mode} & \colhead{Exptime}}
\startdata
A & 4674 & 2004-02-07 & 53042 & Faint\tablenotemark{a} & 18.60\\
B & 7592 & 2007-08-27 & 54339 & VF & 77.09\\
C1& 15595 & 2012-11-28 & 56259 & VF & 62.05\\
C2& 14691 & 2012-12-01 & 56262 & VF & 36.51\\
D & 14692 & 2013-01-25 & 56317 & VF & 103.68\\
E1& 15623 & 2013-03-19 & 56370 & VF & 23.75\\
E2& 15622 & 2013-03-24 & 56375 & VF & 47.04\\
E3& 14693 & 2013-03-27 & 56378 & VF & 22.37\\
F & 14694 & 2013-04-22 & 56404 & VF & 96.24\\
G1& 15551 & 2013-08-25 & 56529 & VF & 30.66\\
G2& 16318 & 2013-08-28 & 56532 & VF & 19.81\\
G3& 16319 & 2013-08-30 & 56535 & VF & 44.48\\
H1& 15552 & 2013-09-16 & 56551 & VF & 36.88\\
H2& 16372 & 2013-09-20 & 56556 & VF & 58.89
\enddata
\tablecomments{ 
The first column indicates observing epochs and how individual exposures were combined for the 2012-2013 monitoring campaign. VF indicates the very faint mode. The last column is the filtered exposure time in ks. }
\tablenotetext{a}{1/8 subarray of the ACIS S3 chip; all other observations used ACIS-I}
\end{deluxetable}

For the spectral data analysis, all data were re-processed with CIAO (version 4.6) utilizing CALDB (version 4.6.3). One event file with the `very faint' correction and one without this correction\footnote{The correction can remove real events in modestly bright sources; see cxc.harvard.edu/ciao/why/aciscleanvf.html} were produced for each observation. We used the former and the latter for the analysis of the extended emission and the  Geminga pulsar, respectively.
All data were checked for flares by filtering the background light curve (0.3-8\,keV, bin 200\,s) for deviations larger than $3\sigma$. The excluded exposure times were usually less than 0.1 ks.
The total exposure for 2004, 2007, 2012 to 2013 is 678\,ks; 
for 2012 to 2013 the total exposure is 582\,ks. On average, each epoch has a combined exposure time of 97\,ks.
To produce exposure-map corrected flux images, the CIAO tasks \texttt{fluximage} and  \texttt{flux\_obs} were used.\\

In order to merge all 2012 and 2013 data, the WCS reference system of exposure F (MJD 56404) was used. Since Geminga has a proper motion of $178.2 \pm 1.8$\,mas\,yr$^{-1}$ \citep{Faherty2007}, the maximal expected shifts of the pulsar were $71$\,mas (F with respect to C) and 74\,mas (F with respect to H). This is on the order of the absolute astrometric uncertainty of {\sl{Chandra}} (90\% uncertainty circle has a radius of $0\farcs{6}$\footnote{http://cxc.harvard.edu/cal/ASPECT/celmon/}).
Position centroids of carefully selected (no counterpart with known proper motion), usually 11 (apparently point) sources in the individual exposures were compared for each exposure with respect to exposure F. Usually, the standard deviation of the shifts in an exposure were larger than or similar to the mean shift found for this exposure, and/or the mean shift was smaller than the {\sl{Chandra}} absolute astrometric uncertainty. In these cases, the respective exposures were not shifted, only re-projected to the reference frame of F.
Two exposures (H1,H2) had relatively large mean shifts (e.g., $0.7\pm 0.4$\,arcsec along one coordinate axis). They were shifted using the task \texttt{wcs\_update}, then re-projected. Exposure G2 (19\,ks) was excluded from merging since there were only few reference sources detected, and the shifts were very different. 
From the remaining 11 exposures, merged event files and exposure-map corrected flux images (energy ranges: broad (0.5-7\,keV), soft (0.5-1.2\,keV), medium (1.2-2\,keV), hard (2-7\,keV)) were produced. 
We use the respective produced merged 2012-2013 event file and exposure-map corrected flux images for defining extraction regions for the extended emission and deriving total count and average flux estimates.\\

The individual exposures of each epoch (labeled by a letter in Table~\ref{obslog}) did not show significant ($>0\farcs{2}$) shifts with respect to each other and were merged on the reference frame of either the longest exposure or the one in the middle of a sequence. We use the respective produced merged event files and exposure-map corrected flux images if we compare the six epochs of 2012-2013.\\

\subsection{Spectral analysis}
Using \texttt{specextract}, we extracted spectra for the pulsar ($r=3\arcsec$) and the extended emission regions outlined in Figure~\ref{Overview} from the individual exposures of 2012/13 and also obtained combined spectra.
The spectra were binned to a minimum of 30 (pulsar) and 25 or 15 (extended emission regions) counts per spectral bin.
We checked different background regions of similar sizes in the north/southeast/west directions of the pulsar (the one in the southeast is shown in Figure~\ref{Overview}). 
Comparing the model fit results for the pulsar as well as for the extended emission, we did not see any significant difference in the fit parameters or fit quality when different background regions were used. All fit results in this paper are given for the background region indicated in Figure~\ref{Overview}.\\

We used Xspec (version 12.8.1g) for the spectral analysis, and applied the Tuebingen-Boulder ISM absorption model (\texttt{tbabs}) with the solar abundance table from \citet{Wilms2000}. \texttt{tbabs} uses the photoelectric cross-section table from \citet{Balu1992} together with He cross-section based on \citet{Yan1998}.
All listed parameter uncertainties indicate the 68\% confidence levels if not otherwise noted.\\

\begin{deluxetable*}{lccc}
\tablecaption{Fit results for the pulsar spectrum in comparison to \citet{deluca2005}\label{pulsartab}}
\tablewidth{0pt}
\tablehead{
\colhead{} & \colhead{multi-fit} & \colhead{combi-fit} & \colhead{De Luca et al. 2005}}
\startdata
instrument & {\sl Chandra} ACIS-I & {\sl Chandra} ACIS-I & XMM-{\sl Newton} EPIC-pn\\
counts     & 19,627 (0.3-8\,keV)   & 20,086 (0.3-8\,keV)               & 52,850 (0.15-8\,keV)\\
net count \%              & $\approx 99.9$	& $\approx 99.9$   & $\approx 95$\\
$N_{\rm H}$ (cm$^{-2}$) & $1.1 \times 10^{20}$ (fixed)   & $1.1 \times 10^{20}$ (fixed) & $1.07 \times 10^{20}$ (fixed)\\
$kT_1$ (eV)          & $59 \pm 3$   & $54 \pm 3$	& $43.1 \pm 0.9$\\
$\mathcal{N}_{\rm BB1}$   & $1.3^{+0.7}_{-0.4} \times 10^4$  & $2.9^{+1.6}_{-1.0} \times 10^4$  & \\

$R_{\rm Em, 1}$ (km)              & $2.9^{+0.7}_{-0.5}$   & $4.2^{+1.0}_{-0.8}$              & $13.7\pm 1.6$\tablenotemark{a}\\
$L^{\rm bol}_1$ ($10^{31}$\,erg/s) & $1.2^{+1.1}_{-0.6}$  & $1.9^{+1.7}_{-0.9}$        & $8.1$\tablenotemark{a}\\

\\
$kT_2$ (keV)               & $0.40 \pm 0.02 $    & $0.40 \pm 0.02$	   & $0.16 \pm {0.03}$\\
$\mathcal{N}_{\rm BB2}$   & $0.25 \pm 0.06 $  & $0.23^{+0.07}_{-0.06}$  	   & \\

$R_{\rm Em, 2}$ (m)             & $12.4^{+1.5}_{-1.6}$  & $11.9^{+1.6}_{-1.7}$       & $64 \pm 16$\tablenotemark{a}\\
$L^{\rm bol}_2$ ($10^{29}$\,erg/s)  &  $5.1^{+2.6}_{-1.8}$ &   $4.6^{+2.6}_{-1.7}$ & $4.1$\tablenotemark{a}\\

\\
$\Gamma$              & $1.47^{+0.06}_{-0.07}$  & $1.53^{+0.05}_{-0.06}$ 	   & $1.7 \pm 0.1$ \\
$\mathcal{N}_{\rm PL}$    & $4.9\pm 0.5$ & $5.3^{+0.4}_{-0.5}$	   & $6.7\pm 0.7$\\
$\chi^2$              & 0.97	& 1.20			   & 1.19\\
dof                   & 559	& 260			   & 73\\

$F_{-13}$ (0.3-8\,keV) & $7.51^{+0.07}_{-0.75}$ & $8.24^{+0.08}_{-0.73}$  \\
$F_{-13}$ (0.2-8\,keV)&   &  & $23$ \\
$F_{-13}$ (1-8\,keV)   & $3.45^{+0.02}_{-0.07}$  & $3.45^{+0.02}_{-0.07}$   & \\
\enddata
\tablecomments{Spectral counts from the 12 Geminga observations in 2012 and 2013 (582,400\,s) are used for our fit. 
The ``multi-fit'' results were derived from a simultaneous fit of the 12 individual spectra, each binned to $N \geq 30$\,counts per bin.
The ``combi-fit'' results were derived from a fit of one spectrum (also binned to $N \geq 30$\,counts per bin) which was obtained as the combination of the 12 individual spectra.
$\mathcal{N}_{\rm BB}$ is in units of km$^{2}$\,(10 kpc)$^{-2}$.
$R_{\rm Em}$ is the radius of an equivalent sphere with the same emission area as the model blackbody. $\mathcal{N}_{\rm PL}$ is in units of $10^{-5}$\,photons\,keV$^{-1}$\,cm$^{-2}$\,s$^{-1}$. $F_{-13}$ are absorbed fluxes in units $10^{-13}$\,erg\,cm$^{-2}$\,s$^{-1}$. 
All listed uncertainties for our fits indicate the 68\% confidence levels for one parameter of interest. 
The distance error is not taken into account in radius and luminosity uncertainties. The conservative luminosity uncertainties were determined from the respective 68\% confidence levels of normalization and temperature.
Note that the cited uncertainties from \citet{deluca2005} are 90\% confidence levels.}
\tablenotetext{a}{re-scaled from $d=157$\,pc to $d=250$\,pc  }
\end{deluxetable*}

\begin{figure*}
\noindent\begin{minipage}[b]{.5\textwidth}
\begin{center}
\includegraphics[height=85mm, angle=-90]{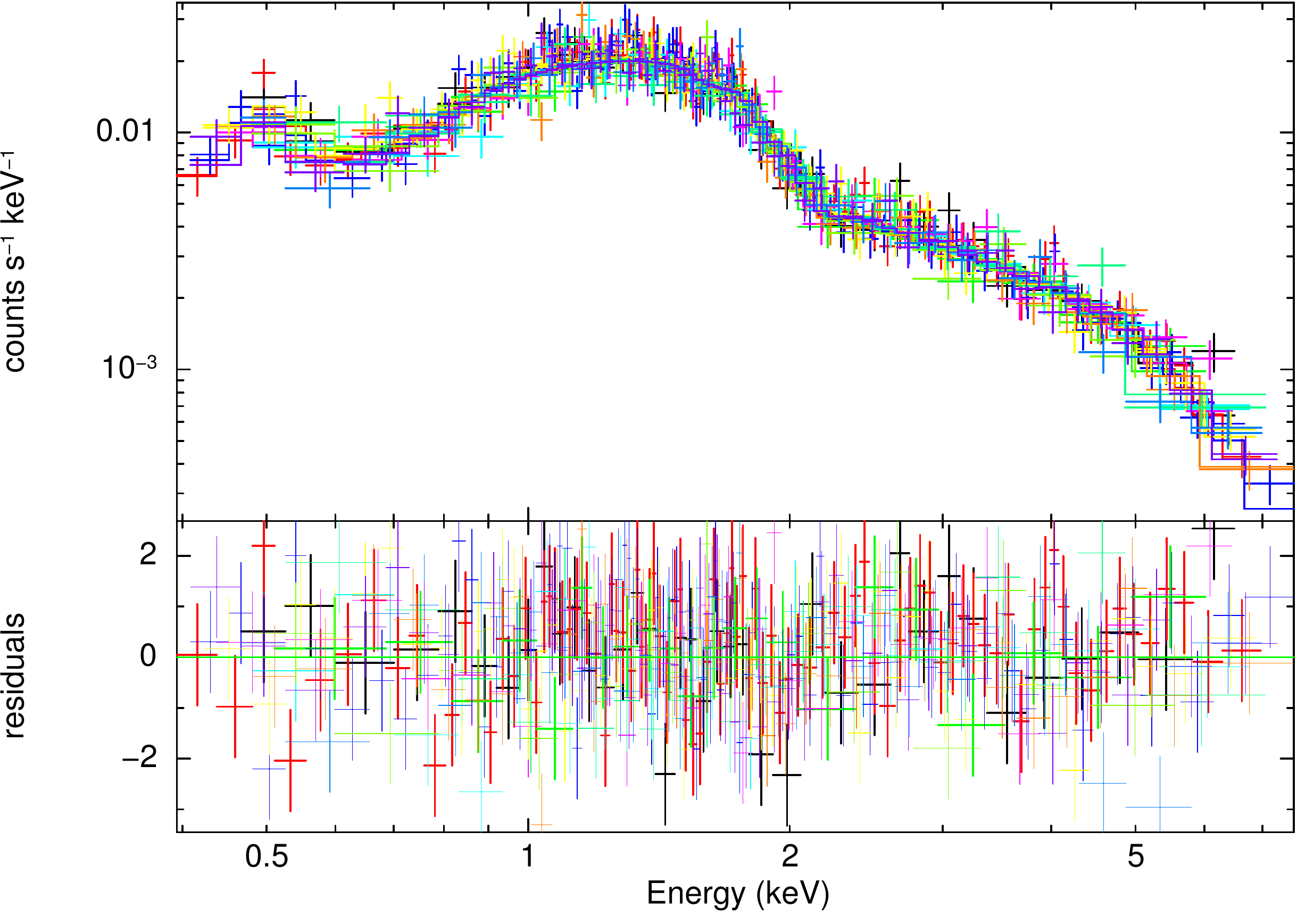}
\end{center}
\end{minipage} 
\begin{minipage}[b]{.5\textwidth}
\begin{center}
\includegraphics[height=85mm, angle=-90]{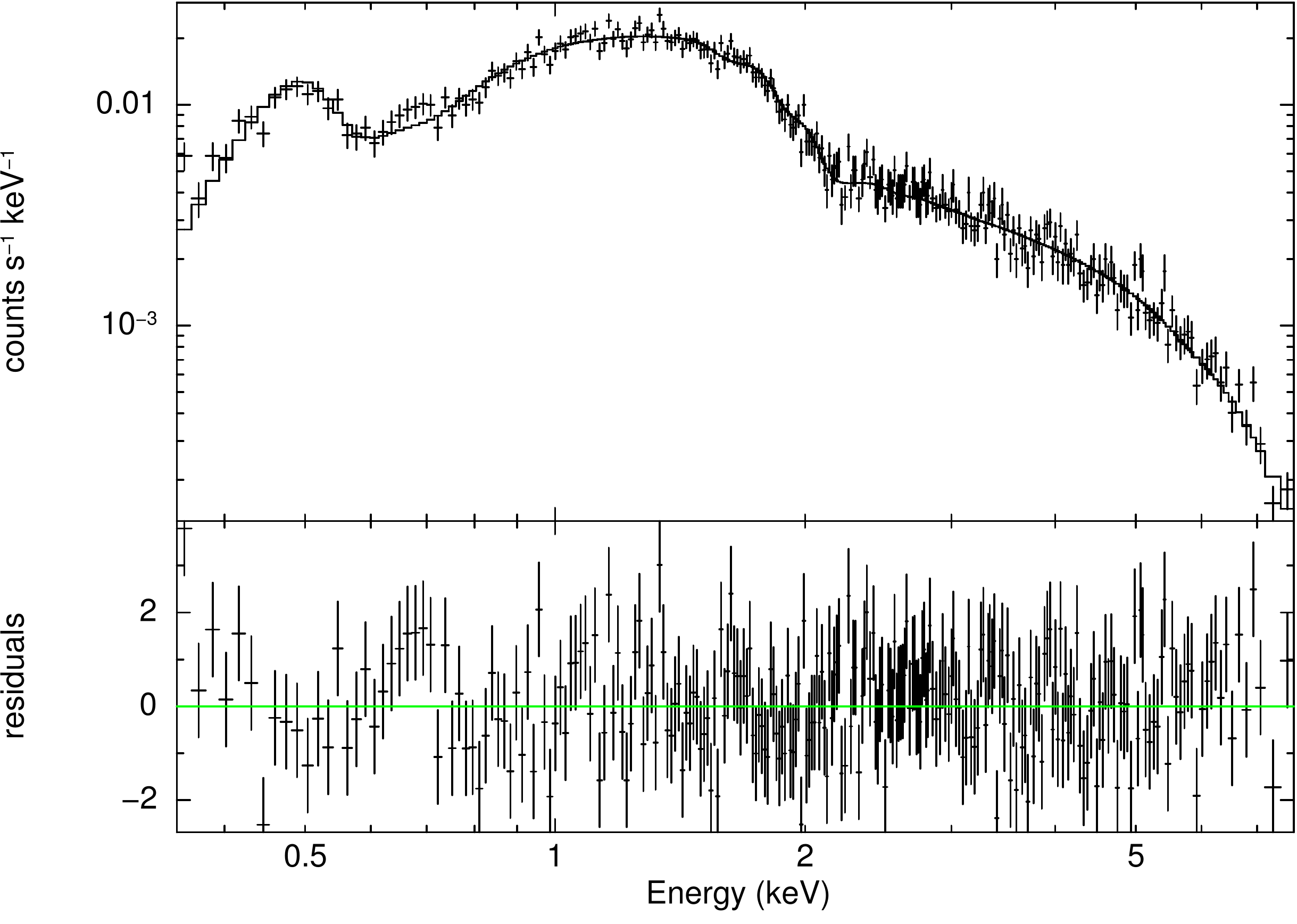}
\end{center}
\end{minipage}
\caption{The three-component spectral model (Table~\ref{pulsartab}) multi-fit {\sl(left panel)} and combi-fit {\sl(right panel)} of the pulsar spectra (12 observations in 2012-2013), binned with $N \geq 30$. Residuals are given in units of $\sigma$. \vspace{0.2cm}}
\label{psrfit}
\end{figure*}

\subsubsection{The Geminga pulsar}
\label{pulsardis}
The detailed physical modeling of the Geminga pulsar spectrum deserves a comprehensive consideration. Since the main focus of this paper is the pulsar wind nebula, we restrict the analysis of the pulsar spectrum to a comparison with previous results, in particular, with those reported by \citet{deluca2005} for dedicated XMM-{\sl Newton} observations.\\

While the many counts of the pulsar allow to consider the spectrum for each epoch individually for spectral fits, the low-count spectra of the extended emission need to be combined.
Aiming to assess the systematic effects of combining individual spectra, we fit the pulsar spectra in two ways. First, in the more rigorous approach we fit all individual 12 pulsar spectra simultaneously -- the `multi-fit' in Table~\ref{pulsartab}. Second, we fit the combined spectrum -- the `combi-fit' in Table~\ref{pulsartab}.  
Similarly to \citet{deluca2005}, we used an absorbed combination of thermal (blackbody (BB)) emission and non-thermal (power law (PL)) emission: $tbabs \times ({\rm BB1}+{\rm BB2}+{\rm PL})$.\\ 

We could only derive an upper limit on $N_{\rm H}< 4.2 \times 10^{20}$\,cm$^{-2}$ from the pulsar multi-fit.
Hence, we fixed its value to the ROSAT result $N_{\rm H}=1.1 \times 10^{20}$\,cm$^{-2}$ \citep{Halpern1997}\footnote{Note that $N_{\rm H}$ is very low and slight changes do not significantly influence our fit results}, the same value that was also applied by  \citet{deluca2005} and that is still within $1\sigma$ error of the latest XMM-$Newton$-NuSTAR study by \citet{Mori2014}. 
The fit results in Table~\ref{pulsartab} indicate an apparently harder pulsar spectrum than seen with XMM (smaller photon index; higher temperatures, in particular for the hot BB component). 
Considering the different extraction region sizes ($45\arcsec$ for XMM-{\sl Newton} and $3\arcsec$ for {\sl Chandra}), one would have expected the reverse result, i.e., that {\sl Chandra} sees less of the hard PWN, thus a softer spectrum. 
On the other hand, {\sl Chandra} is much less sensitive at low energies than XMM-{\sl Newton}, the ACIS response starting to be significant only at 0.3\,keV. Thus, there is less sensitivity to the low blackbody temperatures.
As an explanation for the observed differences, we suspect a (cross-)calibration issue, which could be related, for example, to the time-variable ACIS filter contamination. We  defer a detailed analysis of the pulsar spectrum to another paper.\\

The results of the combi-fit and multi-fit in Table~\ref{pulsartab} are slightly different, the former being slightly softer than the latter. However, values are consistent with each other within their 68\% confidence levels. The reduced $\chi^2$ is worse for the combi-fit and so are the systematic residuals as can be seen from the comparison in Figure~\ref{psrfit}. In particular, there is a strong residual at about $0.5-0.7$\,keV in the combi-fit which is not prominent in the multi-fit.
Since the combi-fit is the less rigorous spectral fit approach, it remains currently unclear whether the feature is real or not. 
It is interesting to note that the combi-fit parameter values can differ by $2\sigma$ from the values obtained with the more rigorous multi-fit approach, even in such a high-count number case.
For the extended emission we only analyze combined spectra due to low count numbers.

\subsubsection{The extended emission}
\label{extem}
\begin{deluxetable*}{lcccccccc}[t]
\tablecaption{Spectral fit results of extended emission\label{extfit}}
\tablewidth{0pt}
\tablehead{
\colhead{Region} & \colhead{area} & \colhead{counts} & \colhead{SF\tablenotemark{a}} & \colhead{$\Gamma$} & \colhead{$\mathcal{N}_{\rm PL}$\tablenotemark{b}} & \colhead{FQ\tablenotemark{c}} & \colhead{$F_{-15}$\tablenotemark{d}} & \colhead{$F_{-15}$\tablenotemark{d}} \\
\colhead{} & \colhead{$[$arcsec$^2]$} & \colhead{} & \colhead{$[\%]$} & \colhead{} & \colhead{} & \colhead{} & \colhead{0.3-8\,keV} & \colhead{1-8\,keV} }
\startdata
Ring  & 136 & 428 &  90 & 1.68$\pm 0.11$ & 1.45$\pm 0.12$ & $0.56/14$ & $9.0\pm 0.5$ & $6.8\pm 0.6$\\
Bow  & 297 & 254 &  60 & 1.27$^{+0.22}_{-0.21}$ & 0.43$^{+0.08}_{-0.07}$ & $1.03/8$ & $3.9\pm0.5$ & $3.4^{+0.6}_{-0.5}$\\
N-tail & 3535 & 1983 & 30 & 0.67$\pm 0.12$ & 1.16$\pm 0.15$ & $0.71/71$ & $21.7^{+2.2}_{-2.0}$ & $20.6^{+2.1}_{-2.2}$\\
S-tail & 4207  & 2883 & 42 & 1.04$^{+0.09}_{-0.08}$ & 2.91$\pm 0.22$ & $1.05/99$ & $34.1^{+2.0}_{-2.2}$ & $31.0^{+2.4}_{-2.2}$\\
A-tail  & 351 & 628 & 80 & 1.63$\pm 0.09$ & 1.84$ \pm 0.13$ & $0.71/22$ &  $11.9^{+0.7}_{-0.6}$ & $9.2^{+0.8}_{-0.6}$\\
A1$+$A2$+$A3  & 240 & 504 &  85 & 1.55$\pm 0.09$ & 1.49$^{+0.12}_{-0.11}$ & $1.22/17$ & $10.2^{+0.5}_{-0.6}$ & $8.2^{+0.7}_{-0.6}$\\
A1$+$A2 & 170 & 414 &  87 & 1.44$^{+0.10}_{-0.09}$ & 1.1$\pm 0.10$  & $1.93/14$ & $8.7^{+0.6}_{-0.5}$ & $7.2 \pm 0.6$\\
A3$+$A2 & 142 & 127 & 76 & 2.40$^{+0.35}_{-0.31}$ & 0.76$^{+0.16}_{-0.14}$ & $0.32/3$ & $3.4^{+0.7}_{-0.5}$ & $1.7\pm 0.3$\\
A4$+$A3$+$A2 & 246  & 251 & 67 & 2.08$^{+0.23}_{-0.21}$ & 0.88$^{+0.13}_{-0.12}$ & $1.11/8$ & $4.4^{+0.5}_{-0.4}$ & $2.7^{+0.5}_{-0.4}$\\
A1 & 97 & 341 &  90 & 1.39$^{+0.11}_{-0.10}$ & 0.98$\pm 0.09$  & $1.08/11$ & $7.8 \pm 0.5$ & $6.5\pm 0.6$\\
\hline
 &  &  &   &  &  & cstat &  &  \\
\hline
A1 & 97 & 367 &  90 & 1.48$^{+0.10}_{-0.11}$ & 1.09$\pm 0.09$  & $196.3/208$ & $8.0 \pm 0.4$ & $6.6\pm 0.3$\\
A2 & 73 & 82 &  68 & 1.74$^{+0.24}_{-0.31}$ & $0.24\pm 0.04$  & $60.1/68$ & $1.4^{+0.1}_{-0.2}$ & $1.07^{+0.05}_{-0.09}$\\
A3 & 70 & 89 &  71 & $2.01^{+0.30}_{-0.26}$ & $0.29^{+0.05}_{-0.04}$  & $60.2/72$ & $1.5\pm0.2$ & $0.95^{+0.05}_{-0.07}$\\ 
A4 & 75 & 70 &  62 & $1.74^{+0.36}_{-0.40}$ & $0.16 \pm 0.03$  & $51.3/58$ & $0.96\pm0.06$ & $0.72 \pm 0.03$ \\
\hline
A4 all & 75 & 120 &  70 & $1.57^{+0.27}_{-0.23}$ & $0.29^{+0.05}_{-0.04}$  & $84.4/96$ & $1.99 \pm 0.14$ & $1.58^{+0.08}_{-0.10}$ 
\enddata
\tablecomments{Fit results for the combined ($\approx 582$\,ks) spectra of the extended emission using a PL model with photon index $\Gamma$. $N_{\rm H}$ was fixed at $1.1 \times 10^{20}$\,cm$^{-2}$. All errors indicate 68\% confidence levels. Total counts and net source percentages (i.e., without background counts) are for energies  $0.3-8$\,keV. Note that removal of `bad bins'  decreased the count numbers of A1 in the $N\leq 25$\,counts\,bin$^{-1}$ case in comparison to the $N= 1$\,counts\,bin$^{-1}$ (cstat) case. In contrast to A4, the region 'A4 all' does not exclude the area of a known (optical/NIR) star, see Section~\ref{starcontam} for a detailed discussion.}
\tablenotetext{a}{SF is the net source count fraction after subtracting the background}
\tablenotetext{b}{The PL norm $\mathcal{N}_{\rm PL}$ is in units of $10^{-6}$\,photons\,keV$^{-1}$\,cm$^{-2}$\,s$^{-1}$ at 1\,keV}
\tablenotetext{c}{fit quality (FQ): either reduced $\chi^2$ / degrees of freedom in the upper part of the table, or cstat / degrees of freedom in the lower part of the table}
\tablenotetext{d}{The absorbed fluxes $F_{-15}$ are in units $10^{-15}$\,erg\,cm$^{-2}$\,s$^{-1}$}
\end{deluxetable*}

\begin{figure*}[]
\noindent\begin{minipage}[b]{.5\textwidth}
\begin{center}
\includegraphics[width=81mm]{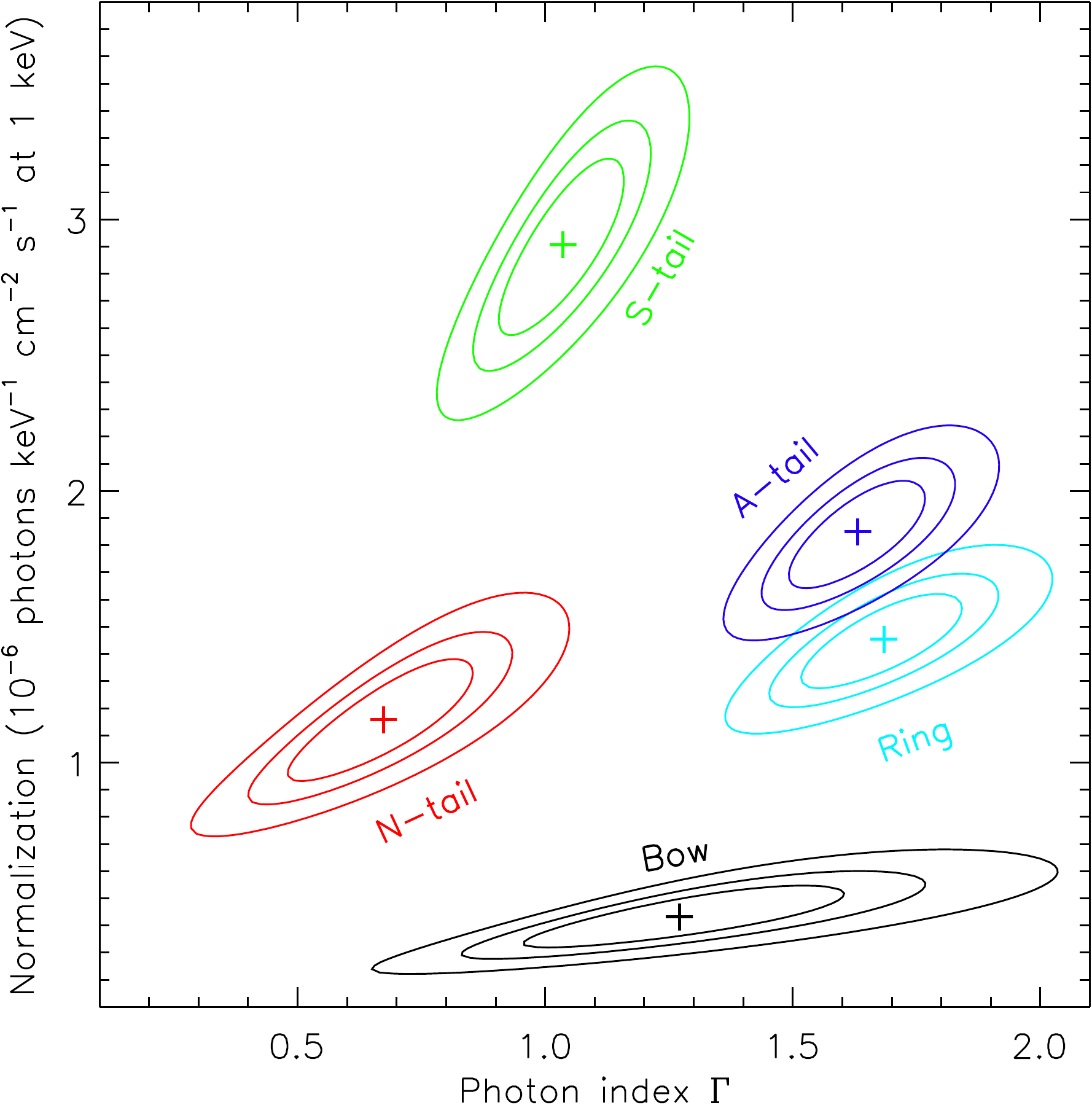}
\end{center}
\end{minipage} 
\begin{minipage}[b]{.5\textwidth}
\begin{center}
\includegraphics[width=85mm]{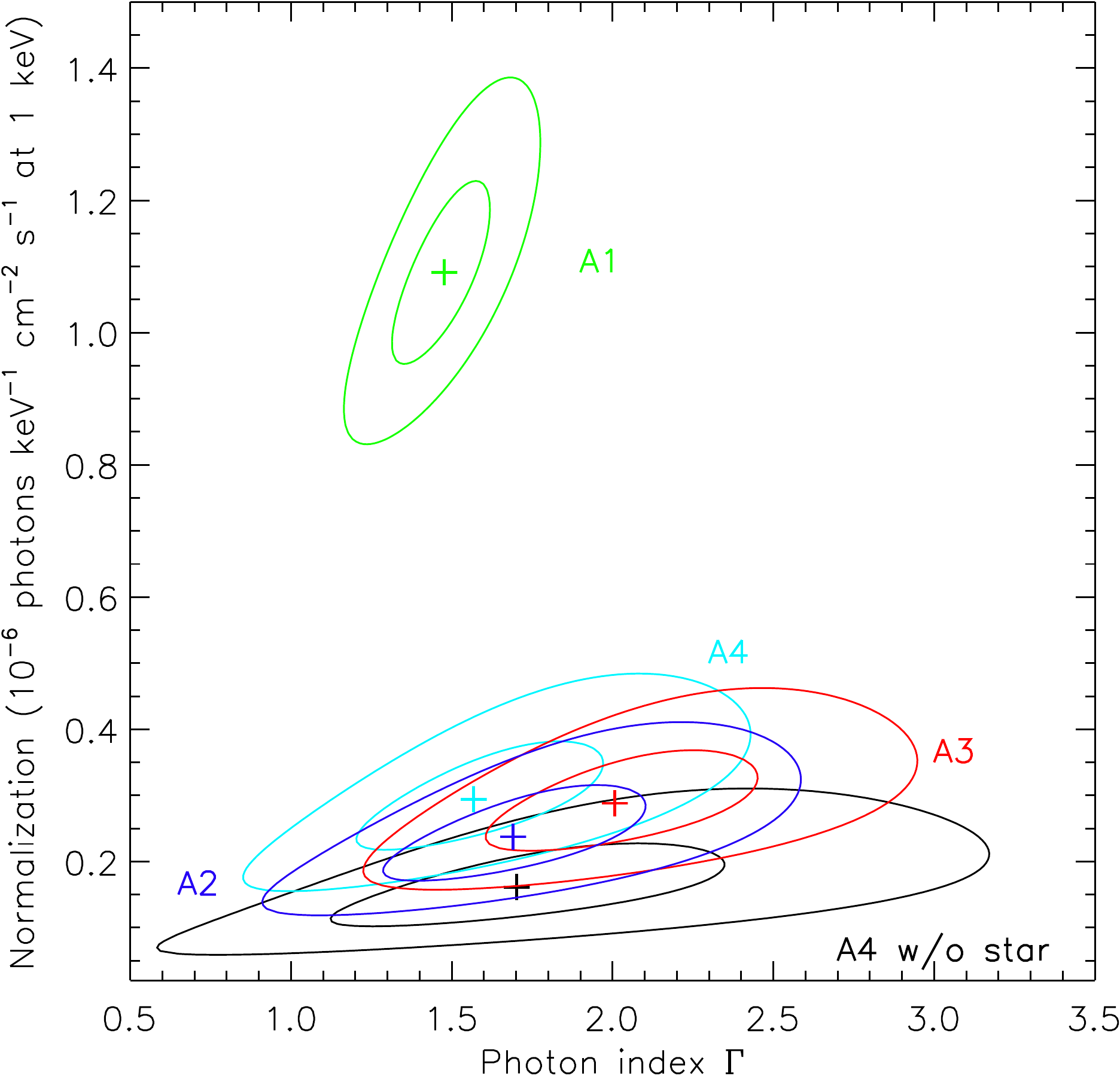}
\end{center}
\end{minipage}
\caption{Confidence contours of the normalization versus the photon index for the PL fit of the axial and lateral tails as labeled in Figure~\ref{Overview}.
\emph{Left panel: } 68\%, 90\% and 99\% confidence levels for the Bow (black), Ring (cyan), N-tail (red), S-tail (green), and the A-tail (blue; the star region in A4 was excluded); obtained from data binned to 25 counts per bin using the $\chi^2$ statistics.   
\emph{Right panel: } 68\%, and 99\% confidence levels for the  A1 (green), A2 ( blue), A3 (red), A4 (black, star region excluded), A4 (cyan; star region \emph{not} excluded); obtained from data with one count per bin using the \texttt{cstat} statistics.}
\label{contours}
\end{figure*}

\begin{figure*}
\noindent\begin{minipage}[b]{.33\textwidth}
\begin{center}
\hspace{1cm}Region: Ring
\includegraphics[width=37mm, angle=-90, clip]{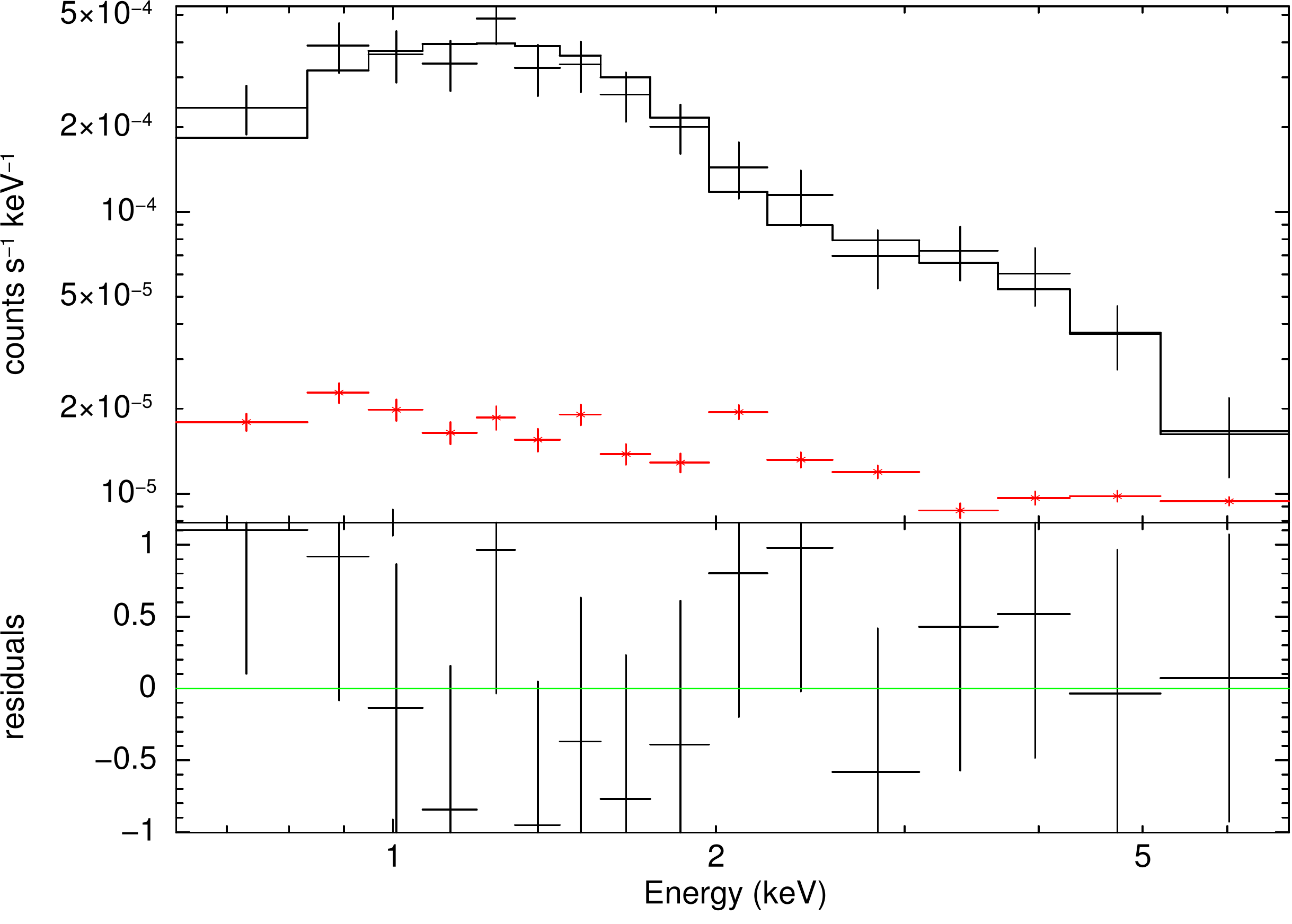}
\end{center}
\end{minipage} 
\begin{minipage}[b]{.33\textwidth}
\begin{center}
\hspace{1cm}Region: Bow
\includegraphics[width=37mm, angle=-90, clip]{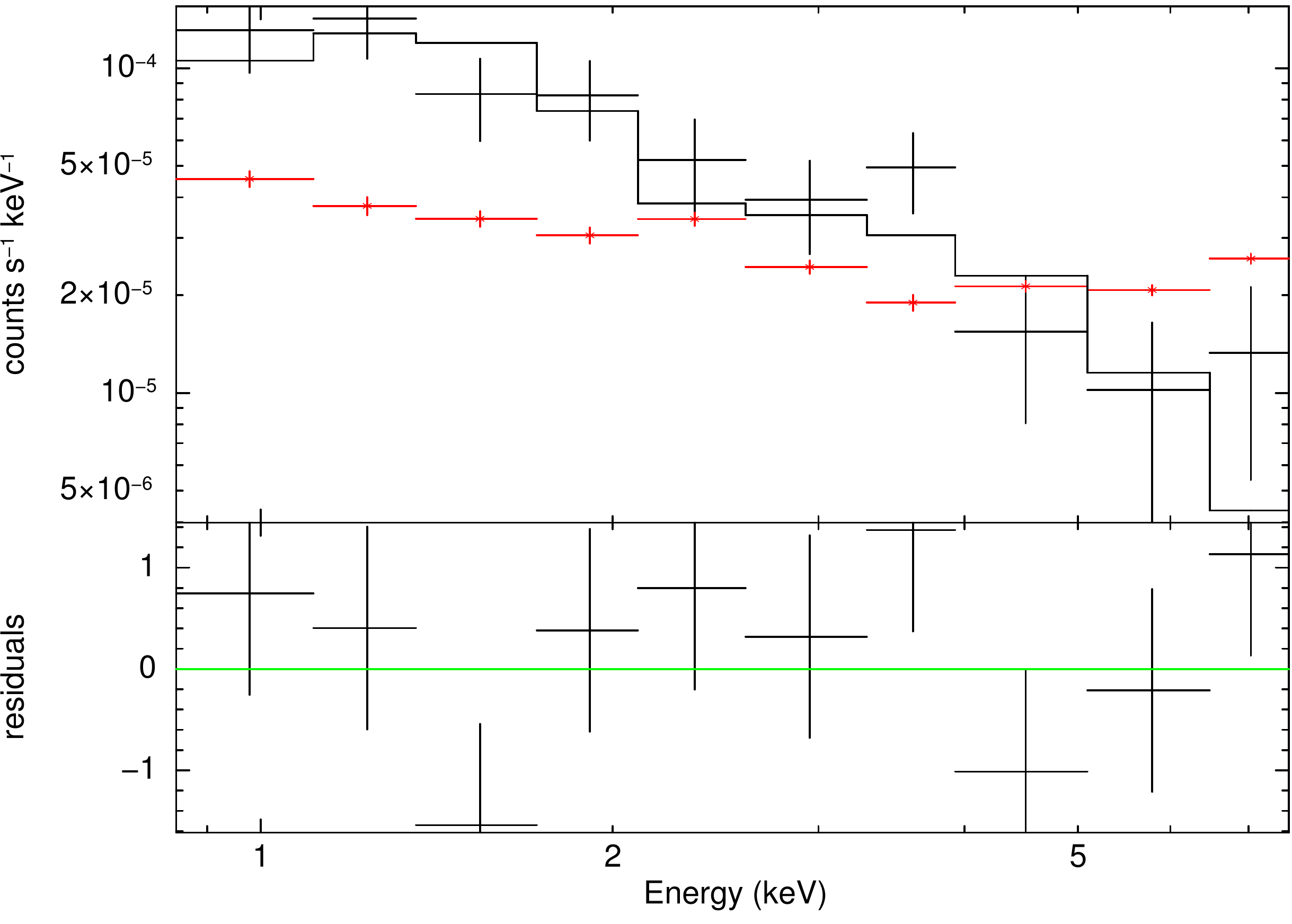}
\end{center}
\end{minipage}
\begin{minipage}[b]{.33\textwidth}
\begin{center}
\hspace{1cm}Region: N-tail
\includegraphics[width=37mm, angle=-90, clip]{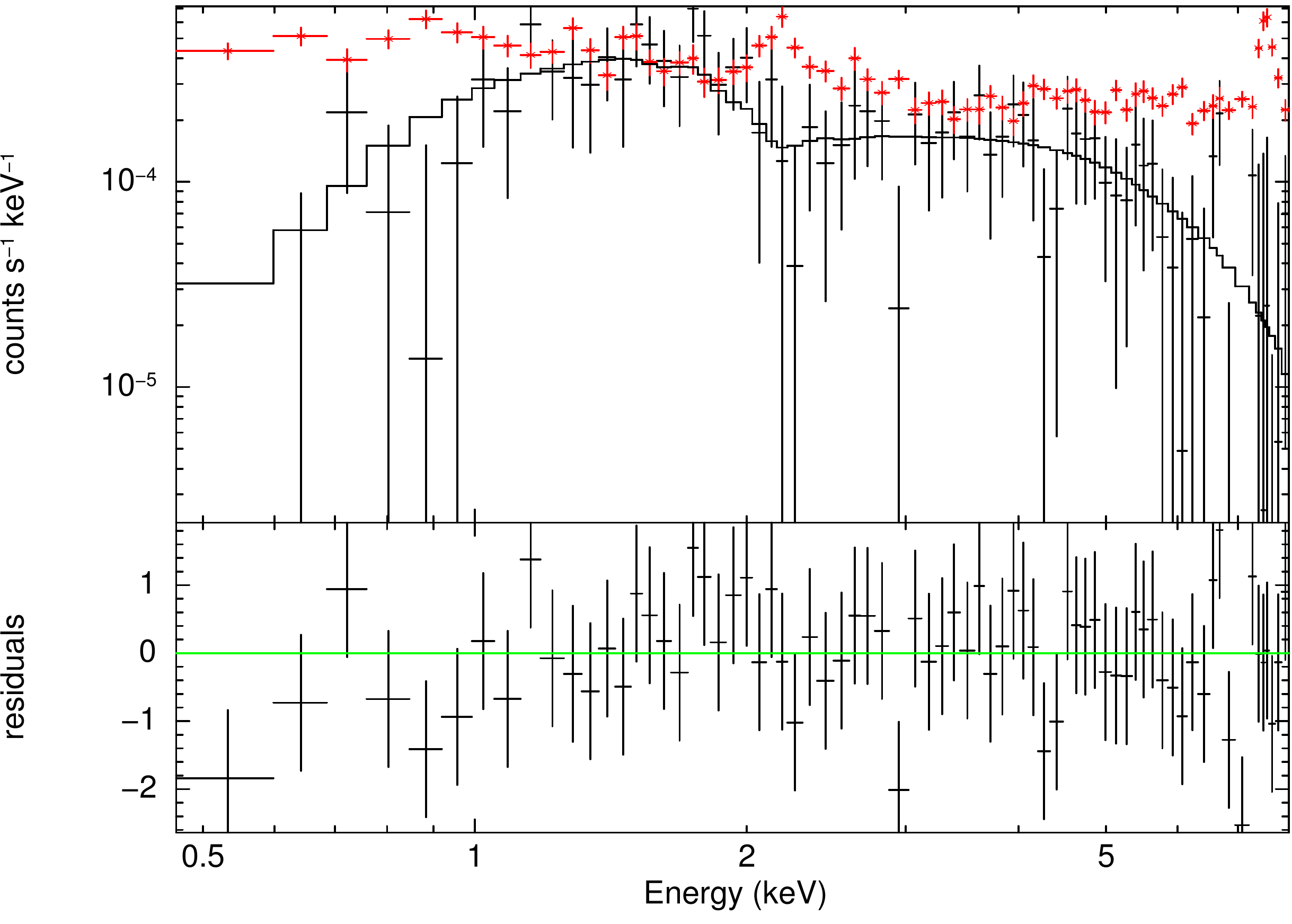}
\end{center}
\end{minipage}
\hfill 
\\ 
\begin{minipage}[b]{.33\textwidth}
\begin{center}
\hspace{1cm}Region: S-tail
\includegraphics[width=37mm, angle=-90, clip]{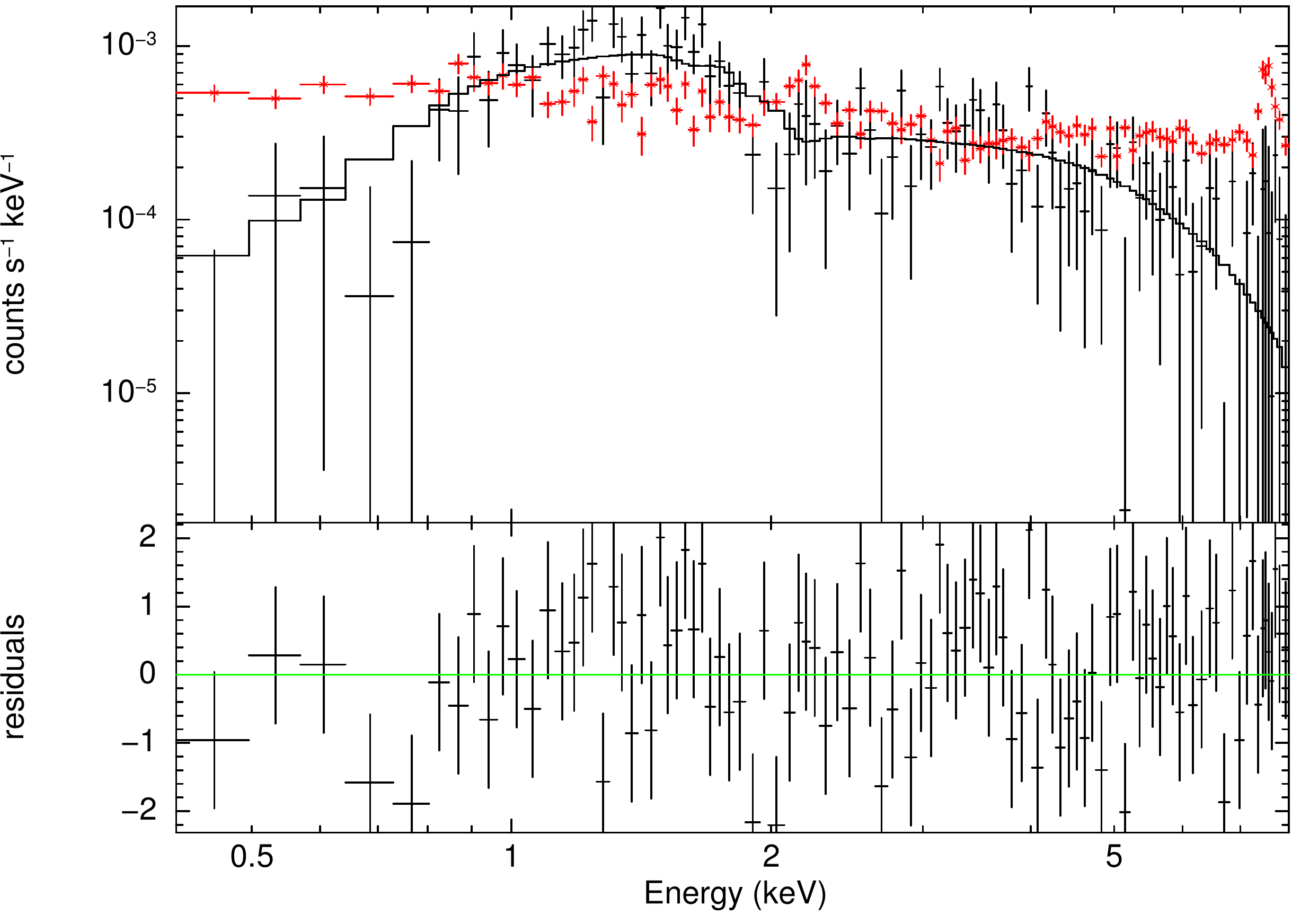}
\end{center}
\end{minipage} 
\begin{minipage}[b]{.33\textwidth}
\begin{center}
\hspace{1cm}Region: A1
\includegraphics[width=37mm, angle=-90, clip]{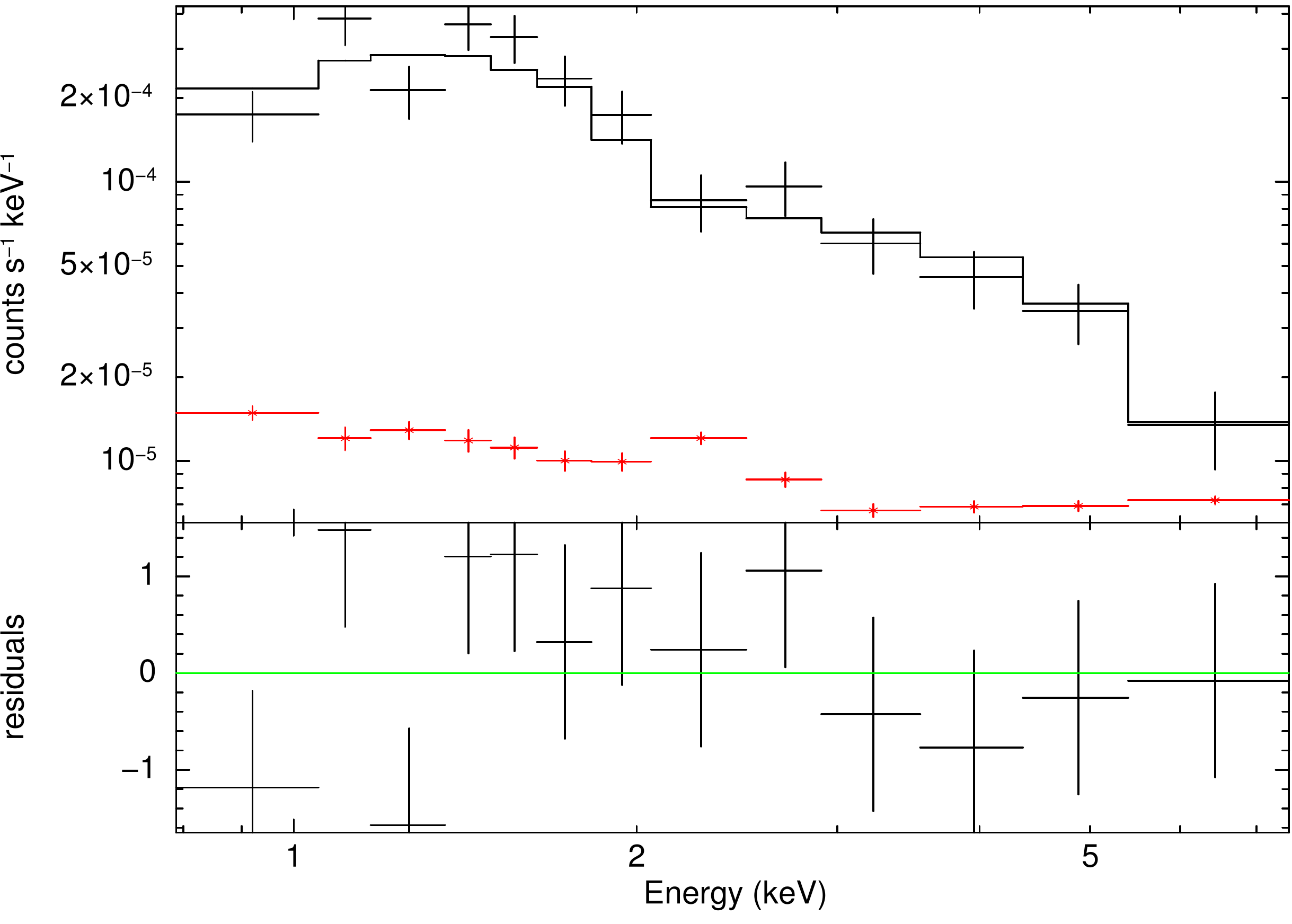}
\end{center}
\end{minipage}
\begin{minipage}[b]{.33\textwidth}
\begin{center}
\hspace{1cm}Region: A1$+$A2
\includegraphics[width=37mm, angle=-90, clip]{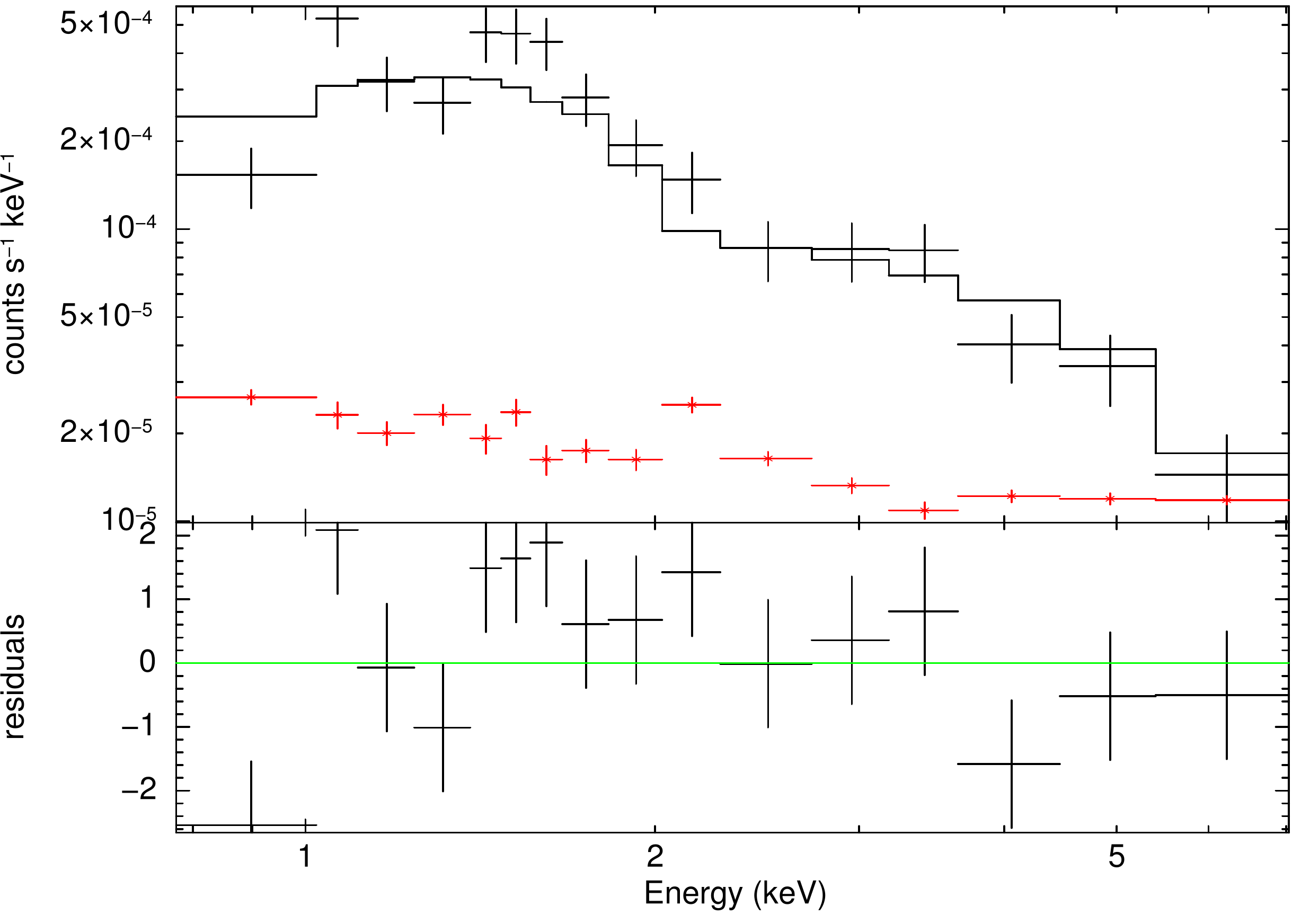}
\end{center}
\end{minipage}
\hfill
\\
\begin{minipage}[b]{.33\textwidth}
\begin{center}
\hspace{1cm}Region: A1$+$A2$+$A3
\includegraphics[width=37mm, angle=-90, clip]{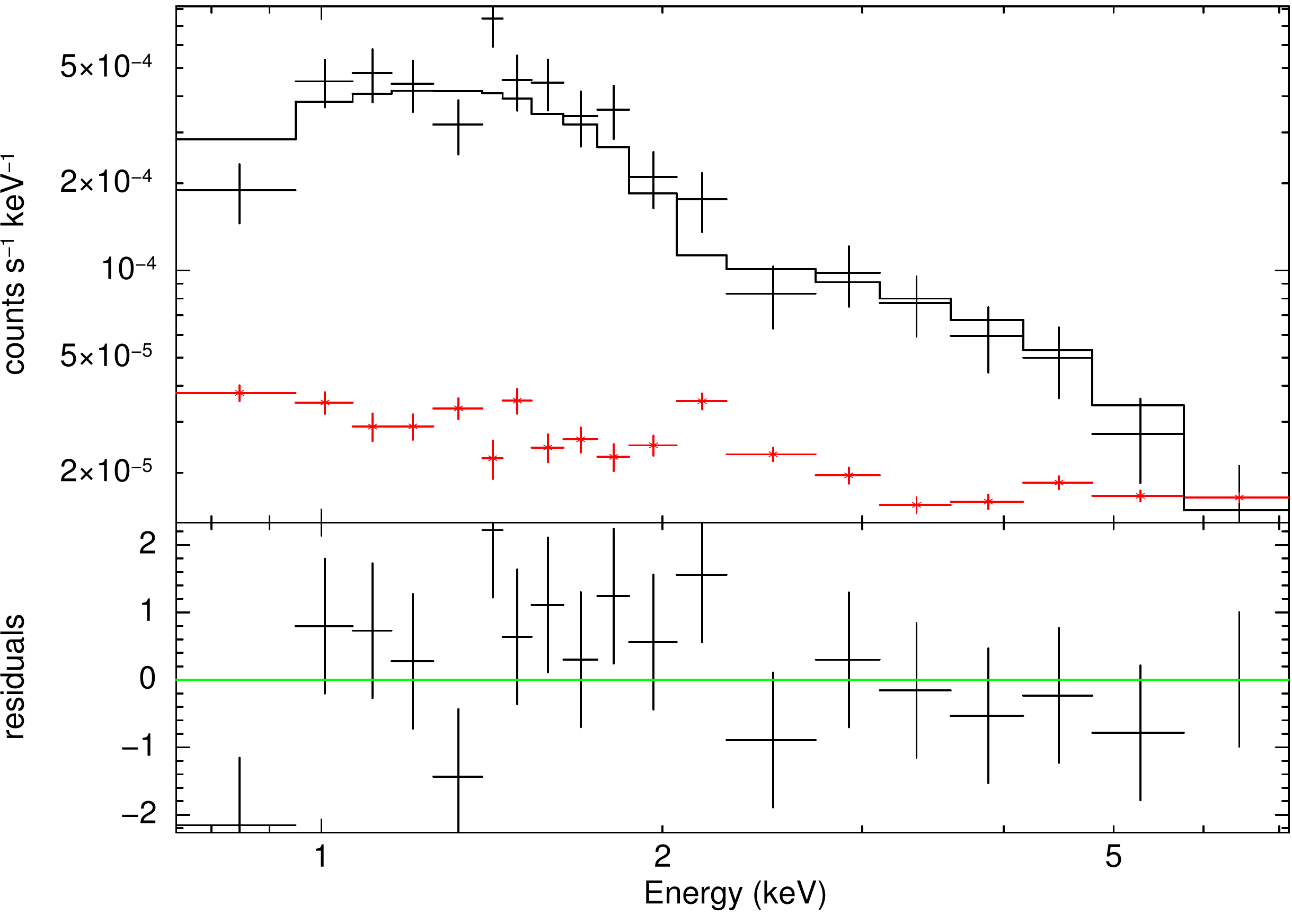}
\end{center}
\end{minipage}
\begin{minipage}[b]{.33\textwidth}
\begin{center}
\hspace{1cm}Region: A3$+$A2
\includegraphics[width=37mm, angle=-90, clip]{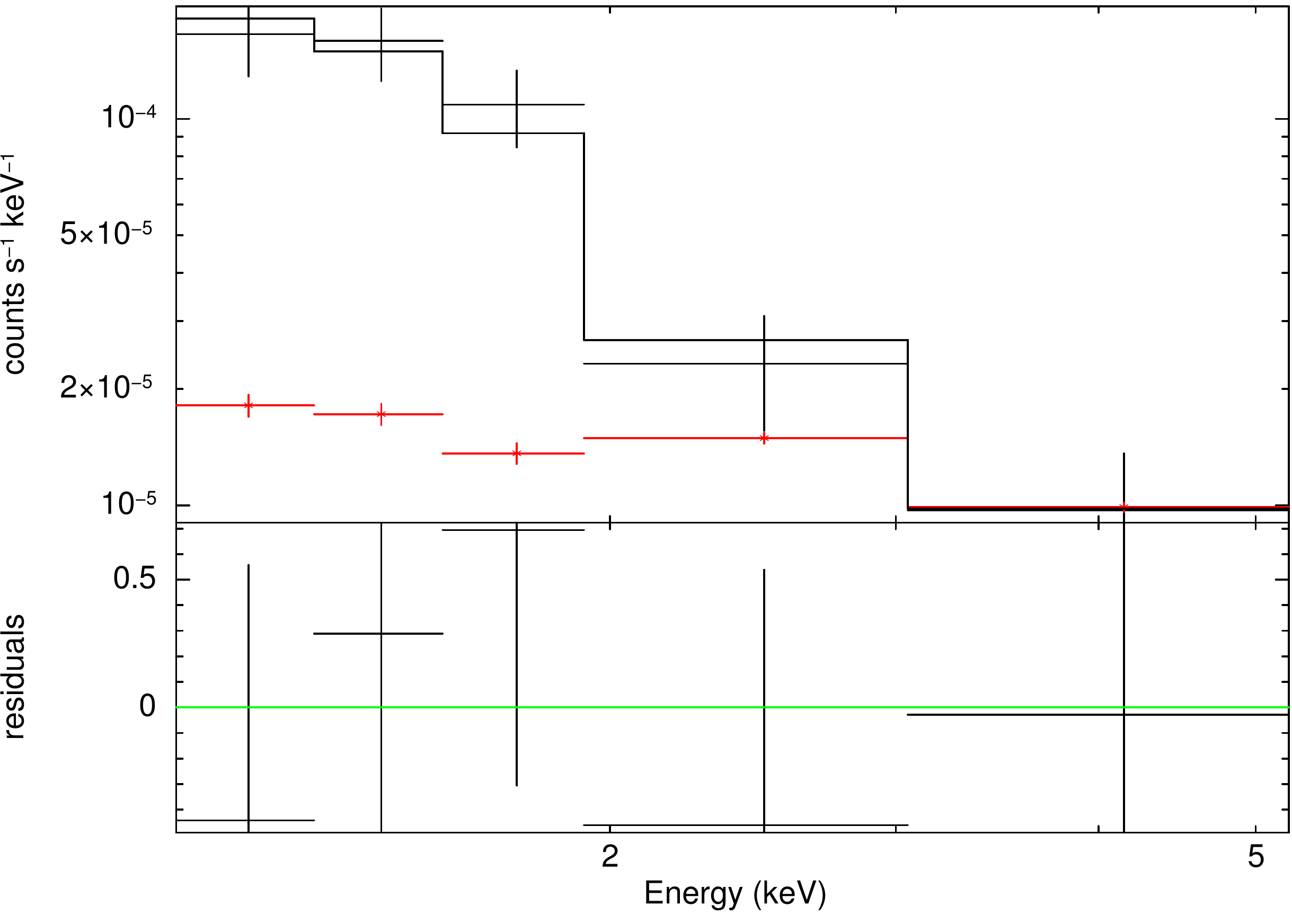}
\end{center}
\end{minipage} 
\begin{minipage}[b]{.33\textwidth}
\begin{center}
\hspace{1cm}Region: A4$+$A3$+$A2
\includegraphics[width=37mm, angle=-90, clip]{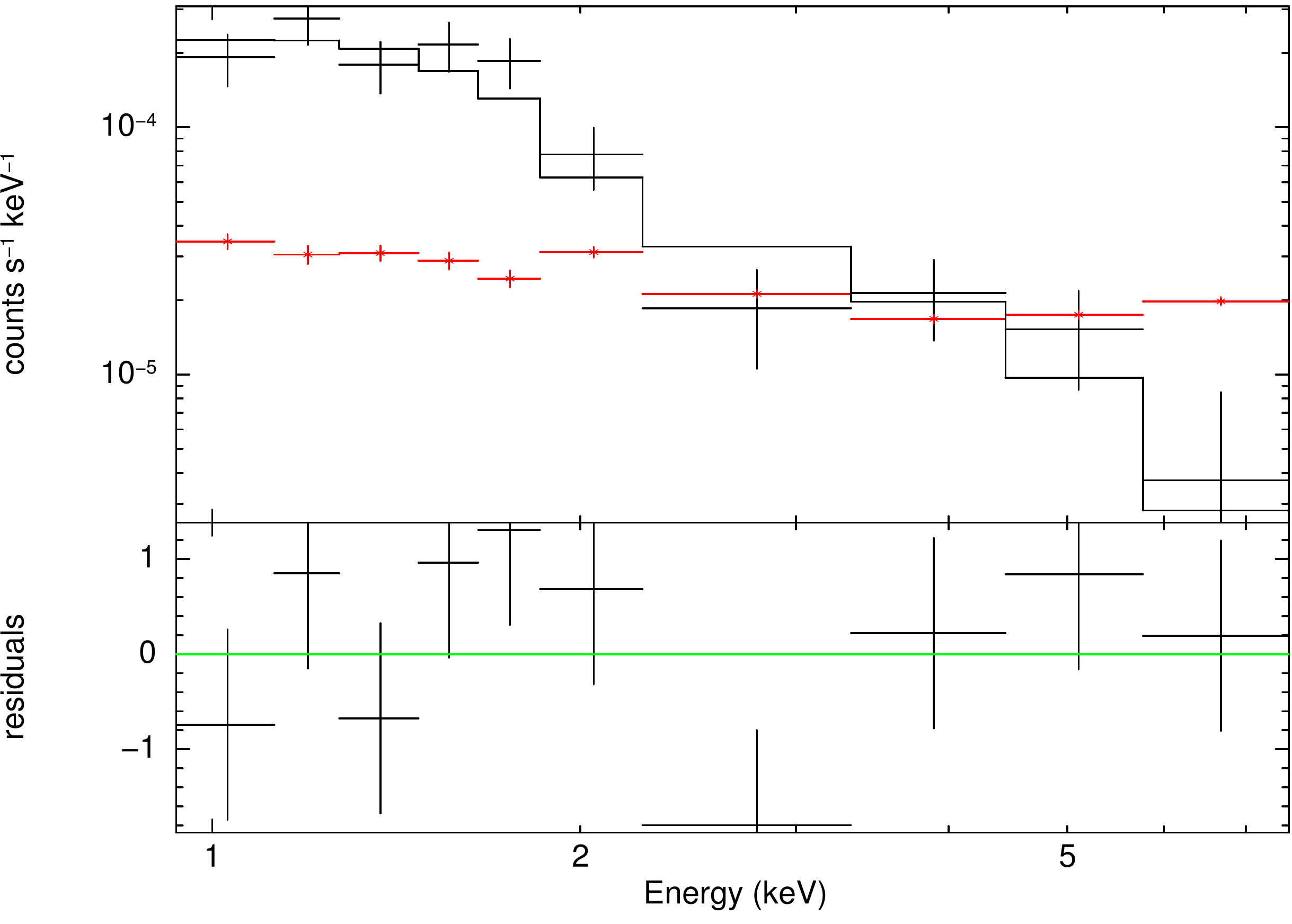}
\end{center}
\end{minipage}
\hfill
\\
\noindent\begin{minipage}[b]{.33\textwidth}
\begin{center}
\hspace{1cm}Region: A-tail
\includegraphics[width=37mm, angle=-90, clip]{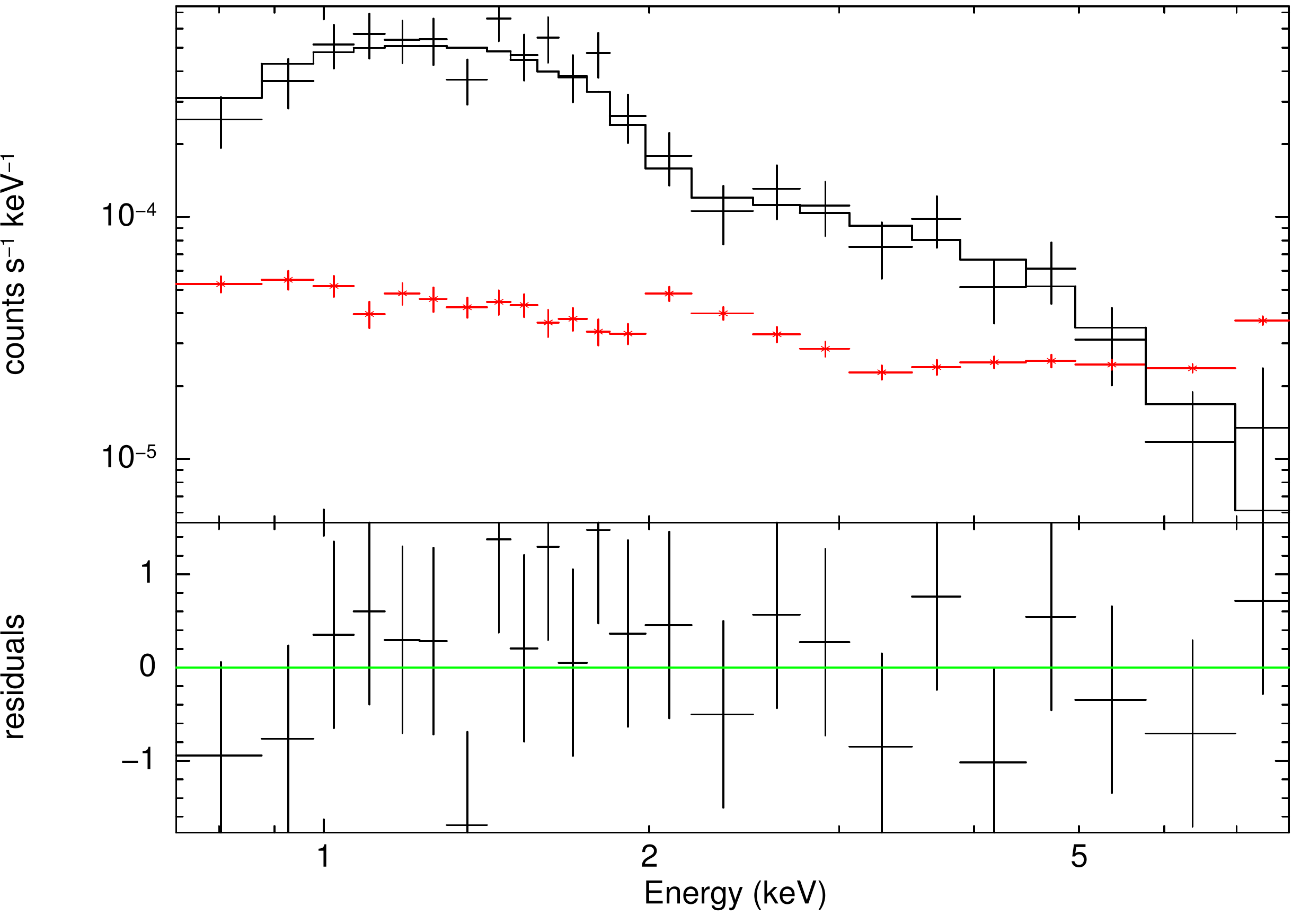}
\end{center}
\end{minipage}
\begin{minipage}[b]{.33\textwidth}
\hfill
\end{minipage}
\begin{minipage}[b]{.33\textwidth}
\begin{center}
\hspace{1cm}Region: A4 (cstat)
\includegraphics[width=37mm, angle=-90, clip]{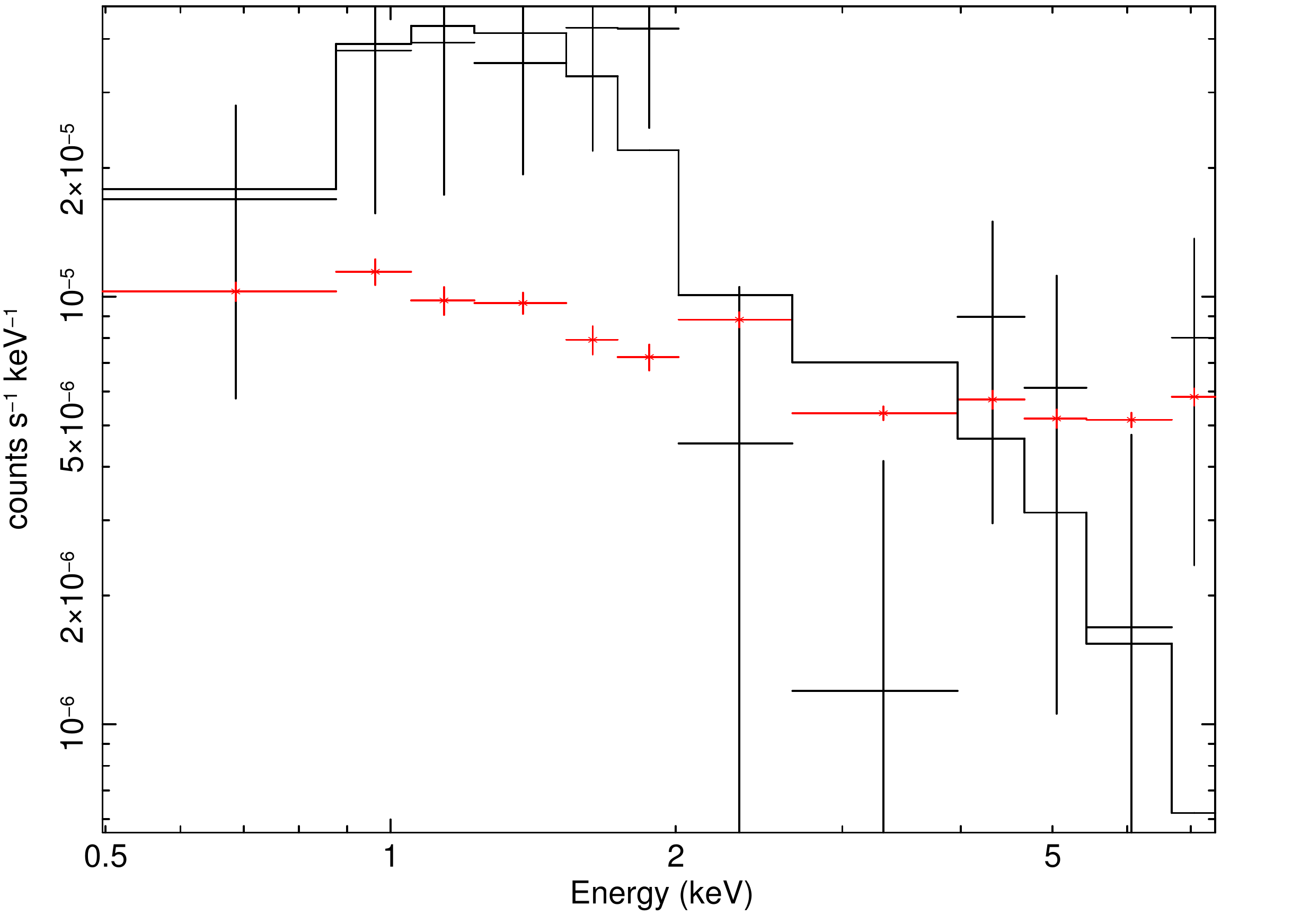}
\end{center}
\end{minipage}
\\
\\
\noindent\begin{minipage}[b]{.33\textwidth}
\begin{center}
\hspace{1cm}Region: A3 (cstat)
\includegraphics[width=37mm, angle=-90, clip]{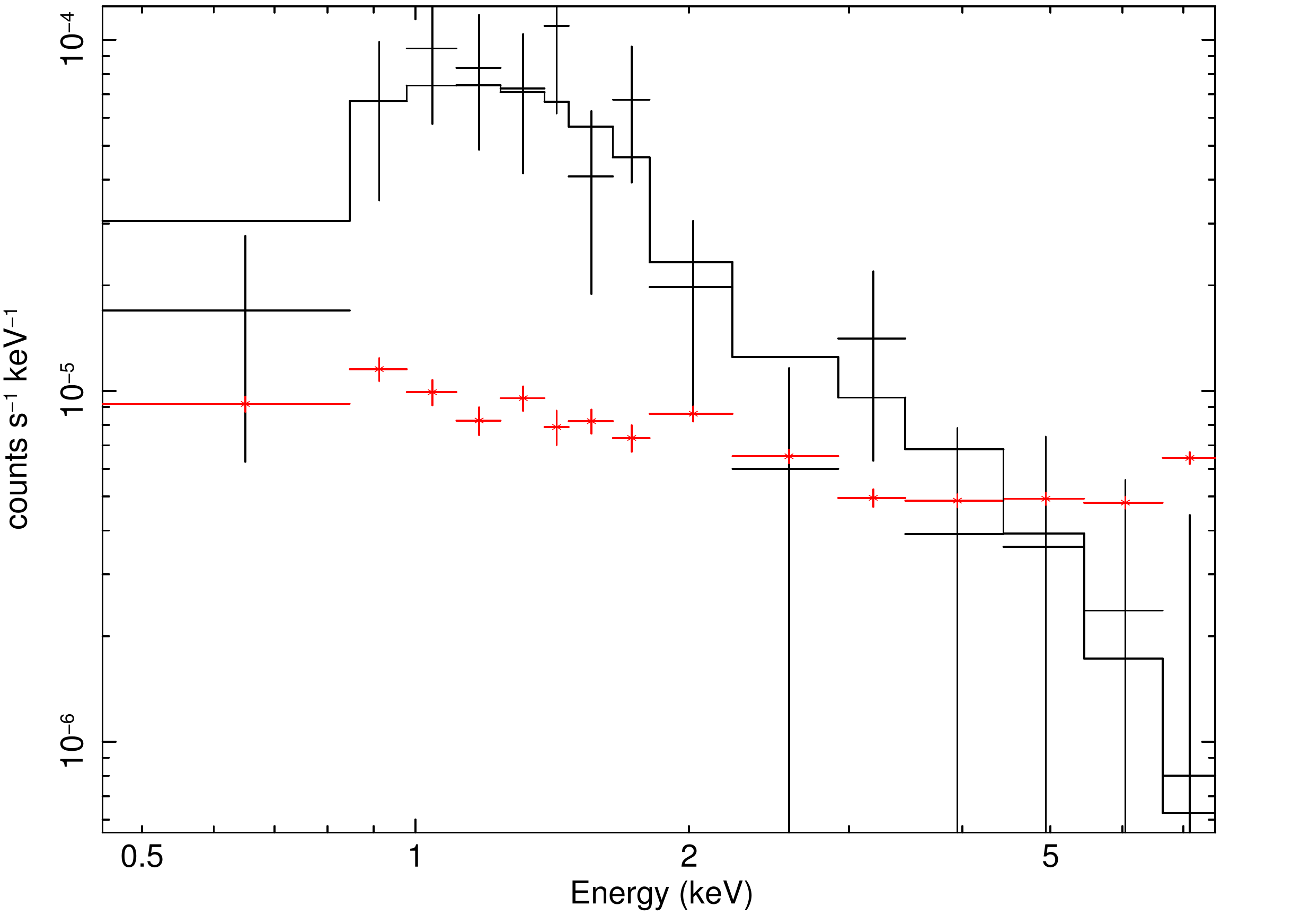}
\end{center}
\end{minipage}
\begin{minipage}[b]{.33\textwidth}
\begin{center}
\hspace{1cm}Region: A2 (cstat)
\includegraphics[width=37mm, angle=-90, clip]{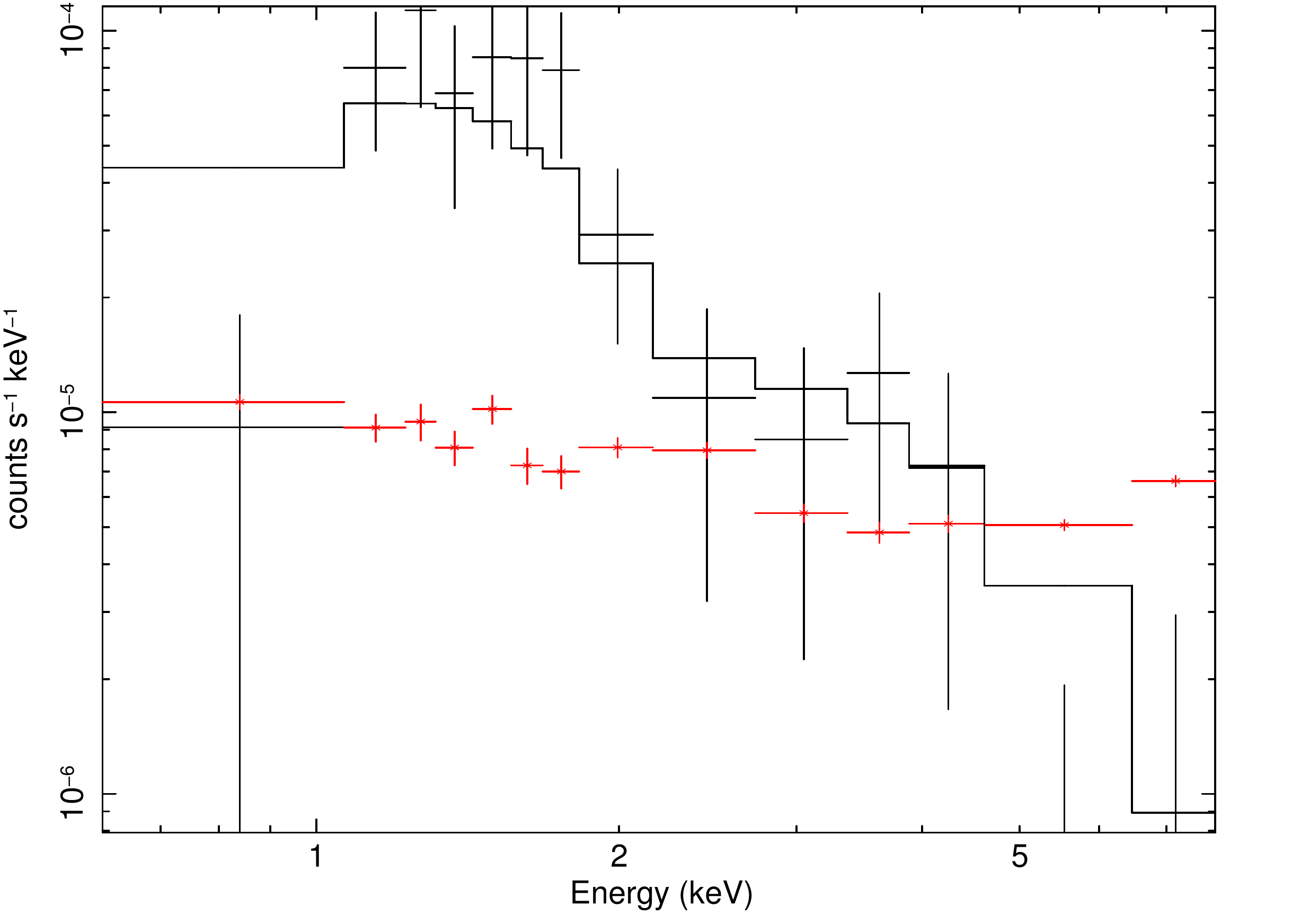}
\end{center}
\end{minipage}
\begin{minipage}[b]{.33\textwidth}
\begin{center}
\hspace{1cm}Region: A1 (cstat)
\includegraphics[width=37mm, angle=-90, clip]{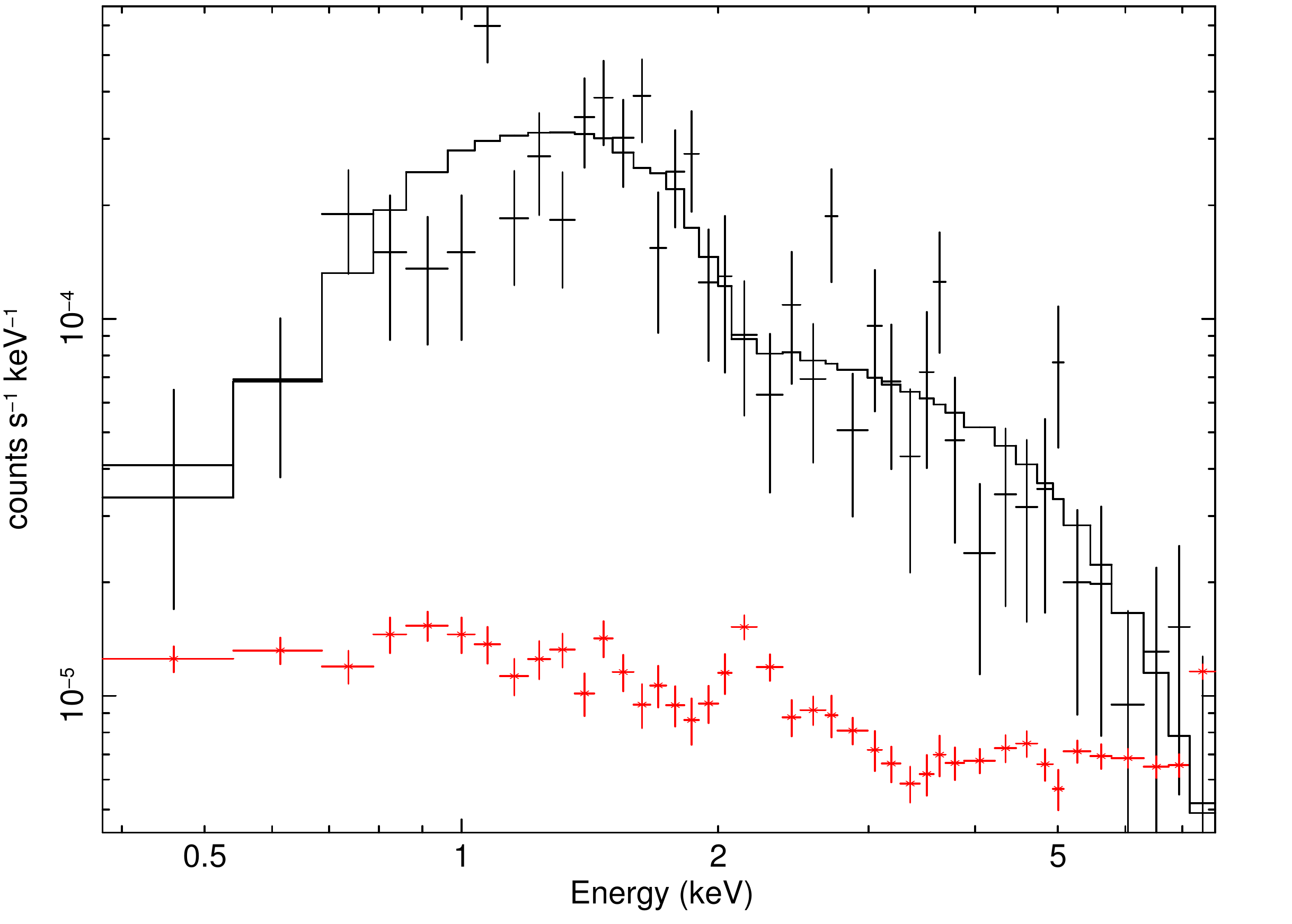}
\end{center}
\end{minipage}
\hfill
\caption{PL fits of the combined spectra (at least 25\,counts\,bin$^{-1}$ for first 13 panels, 5\,counts\,bin$^{-1}$ for last 4 (cstat) panels) of the extended regions defined from the observations of 2012-2013 (Figure~\ref{Overview}). The background-subtracted data points are shown in black, the respective (area-scaled) background points are plotted in red. The model fit parameters correspond to those listed in Table~\ref{extfit}. The residuals for the $\chi^2$-fits are shown in units of $\sigma$.}
\label{PLfits}
\end{figure*}

As described by PBZ10, one can differentiate between different extended emission regions: the `lateral tails' (`N-tail' and `S-tail' in Figure~\ref{Overview}), the `axial tail' (`A1' to `A4' in Figure~\ref{Overview}), and an arclike diffuse emission region in front of the pulsar (the `Ring' and the `Bow' in Figure~\ref{Overview}). 
The bent lateral tails have extensions of about $3\farcm{1}$, and the axial tail has an extension of about $45\arcsec$. We split the axial tail in four individual regions which are detected at different times (see Section~\ref{Atail}). We note that for the spectral analysis, however, we used the combined spectra from all observations from 2012-2013 due to low count numbers. When considering spatial combinations of A1 to A4, we created new regions whose boundaries follow closely the outer boundaries of the individual regions.  
The Ring covers an arc region with $r\approx 3\arcsec - 8\arcsec$ from the pulsar, the Bow region has a width of about $8\arcsec$.\\

We checked optical, near-infrared and infrared data for potentially contaminating sources in our source extraction regions.
We applied circular exclusion regions with radii of $2\arcsec-3\arcsec$ depending upon the location and brightness of the potentially contaminating  source.
There are two cases of clear enhancements in X-ray counts (south tip of the S-tail, near the bend of the N-tail) due to background/foreground sources. Following a conservative approach, we also excluded other regions, where a contribution to the X-ray counts seemed unlikely but could not be entirely ruled out.
For A4, for example (where there is no obviously enhanced X-ray emission at the position of a known star), this can lead to a slight underestimate of the source flux.\\

We fit all extended region spectra that have more than 100 counts with an absorbed PL, where  $N_{\rm H}=1.1 \times 10^{20}$\,cm$^{-2}$ is fixed\footnote{Using $N_{\rm H}=2.34 \times 10^{20}$\,cm$^{-2}$ -- the maximal possible (best-fit value $+3\sigma$) value for a reasonable pulsar fit according to \citet{Mori2014} -- results in negligible changes of the spectral fit parameters.}. 
The derived spectral fit parameters are listed in Table~\ref{extfit}. Spectra, their fits and residuals, and confidence contours are shown in  Figures~\ref{PLfits} and \ref{contours}, respectively.
The most interesting results are the small photon indices ($\Gamma=0.7\pm 0.1$ and $1.0\pm 0.1$) of the lateral tails. This emission is significantly harder than that of the pulsar non-thermal emission ($\Gamma=1.53^{+0.05}_{-0.06}$, combi-fit).
The Ring and the Bow have photon indices which are close to the pulsar's one, with the bow being slightly harder ($\Gamma=1.3\pm 0.2$) than the pulsar.
The photon index of the whole axial tail, A-tail, is close to the value of the pulsar as well. However, count numbers of the A-tail are dominated by A1, the region closest to the pulsar. For the regions further away from the pulsar, A4$+$A3$+$A2 ($\Gamma=2.1\pm 0.2$) and A3$+$A2 ($\Gamma=2.4^{+0.4}_{-0.3}$), the emission apparently becomes softer.
A hint of this trend can already be seen comparing the values for A1, A1$+$A2,  A1$+$A2$+$A3, and the whole A-tail (see Table~\ref{extfit}). 
Using the \texttt{cstat} statistic we checked whether we could verify this trend for the low-count regions A4 to A1 (lower part of Table~\ref{extfit}). The $\Gamma$ difference is largest for A3 ($\Gamma=2.0 \pm 0.3$) and A1 ($\Gamma=1.5 \pm 0.1$). However, even these values overlap within their $2\sigma$ uncertainties. The $\Gamma - \mathcal{N}_{\rm PL}$ confidence contours (Fig.~\ref{contours}) show overlapping 68\% confidence contours for A2 to A4, while A1 is significantly brighter. 
We conclude that though we see a trend towards softening in the outer axial tail, the current count numbers are too low to confirm it with sufficient significance.
We explored dividing the lateral tails in different sections and compared the derived $\Gamma$. There was no significant change in $\Gamma$ over the lengths of the lateral tails.\\

Regarding the lateral N- and S-tail, we investigated whether thermal plasma emission (\texttt{APEC}) or bremsstrahlung could be a reasonable spectral model.
The fits resulted only in lower limits for $kT$.
We obtained $kT>51$\,keV (APEC), $kT>132$\,keV (bremsstrahlung) and $kT>34$\,keV (APEC), $kT>121$\,keV (bremsstrahlung) for the N-tail and the S-tail, respectively. 
These temperatures are unphysically high. Thus, we exclude thermal plasma emission as powering the lateral tails.

\begin{figure*}
\begin{center}
\includegraphics[width=150mm]{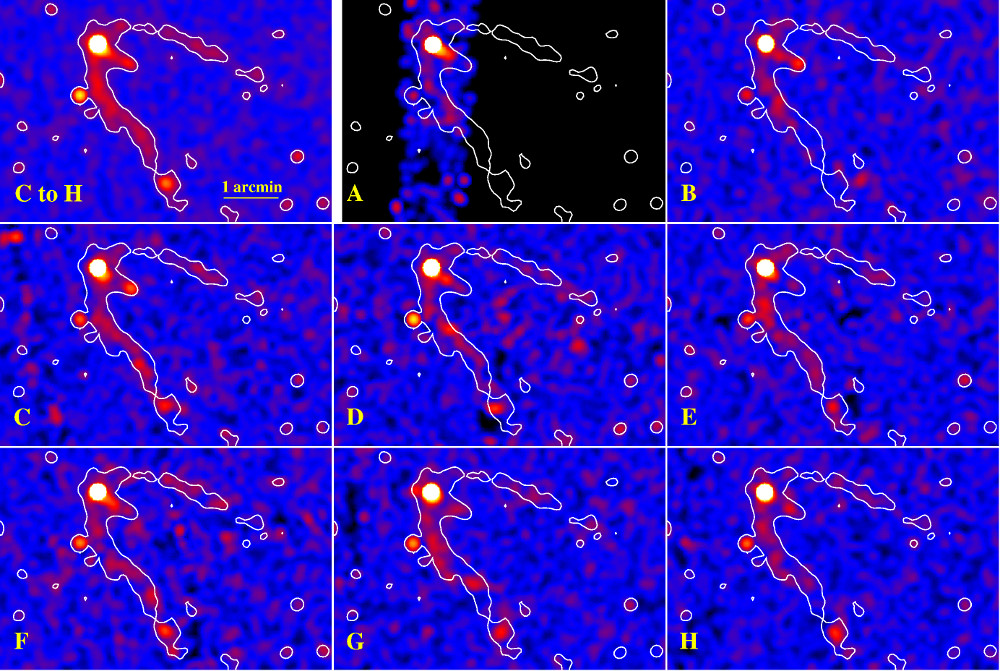}
\includegraphics[width=150mm]{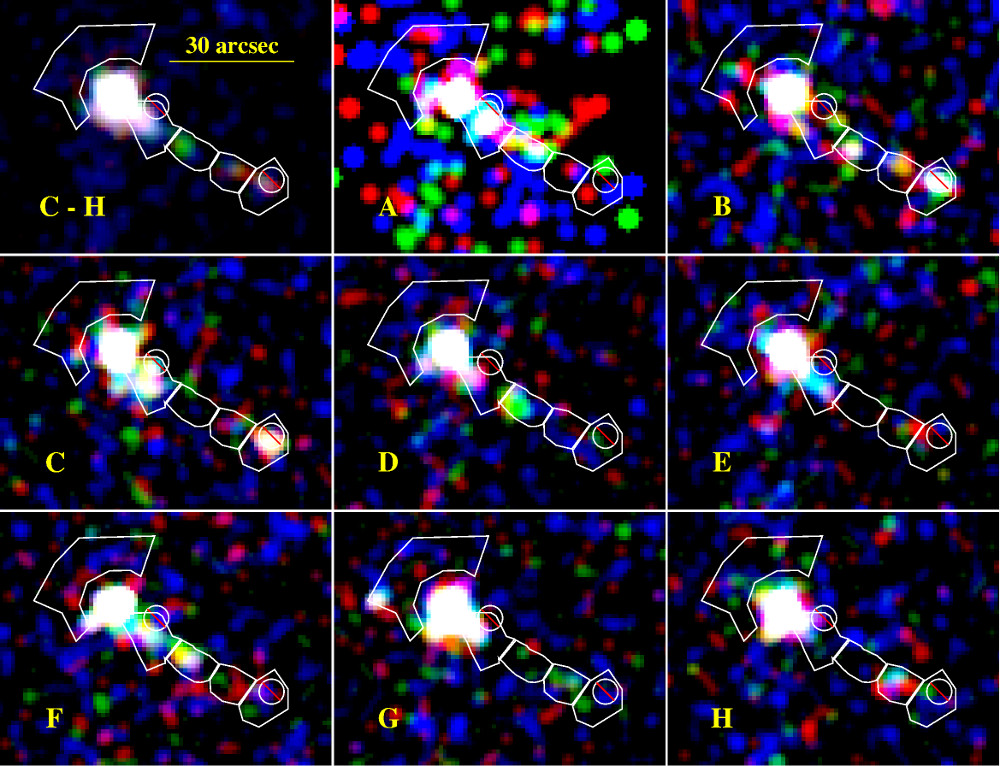}
\end{center}
\caption{\protect\\{\sl{Upper 9 panels:}} Exposure-corrected broad-band $[0.5,8$\,keV$]$ flux images of the Geminga  PWN during the individual observing epochs (see Table~\ref{obslog}). Images have been binned to pixel sizes of $0\farcs{984}$ (2 ACIS pixels) and smoothed using a Gaussian with $\sigma = 9$\,pix. The logarithmic color scale is slightly tuned in each panel to highlight interesting features.   
The merged 2012/13 image in the first panel is employed to obtain a contour level which is also overplotted in all other images to guide the eye. For each epoch, there are \emph{on average} about 170, 90, 70 net counts in the S-, N- and axial tail, respectively. \protect\\
{\sl{Lower 9 panels:}} Exposure-corrected RGB band flux images of Geminga's axial tail during the respective observing epochs. The R, G, B bands correspond to energies of $[0.5,1.2$\,keV$]$,$[1.2,2$\,keV$]$, $[2, 7$\,keV$]$, respectively. Flux images have been binned to pixel sizes of $0\farcs{984}$ (2 ACIS pixel) and smoothed using a Gaussian with $\sigma = 3$\,pix. The logarithmic color scale is the same for all bands and panels. The overplotted regions are A4 to A1 and the Bow region from Figure~\ref{Overview}.}
\label{tempchanges}
\end{figure*}

\subsection{Changes over time}
\label{sec:tempchanges}
Figure~\ref{tempchanges} gives an overview of the temporal changes in the extended emission around Geminga. 
The upper 9 panels emphasize changes on larger scales, i.e., the lateral tails, in the broad energy band ($0.5-8$\,keV); the lower 9 panels are a zoom-in RGB image of the axial tail in the soft ($0.5-1.2$\,keV), medium ($1.2-2.0$\,keV), and hard ($2.0-7$\,keV) energy bands. We remind the reader that the observations in epochs A and B are not as deep as in the later epochs C to H. After a general description of temporal changes we report on our investigation of individual features.\\

Slight brightness changes of the lateral tails seem to be consistent with random fluctuations judging from fluctuations in the background emission and considering the low count numbers of the extended emission.  
In the first part of the S-tail (close to the pulsar) there appears to be a brightness shift by $10\arcsec-15\arcsec$ within the combined outline of the S-tail -- in epoch C, this part has a smaller angle with respect to the axial tail (and a larger separation from the east contour in Figure~\ref{tempchanges}) than in epoch D or F. 
Slight brightness changes are seen at the S-tail's `knee' located at about half of its length. The emission appears to resemble a linear feature in epochs C and D, but starts to bend in E. The `knee' is most pronounced in epoch G.
There is also some indication for strengthening of the emission at the southern tip of the S-tail.
However, due to low count numbers statistical fluctuations cannot be excluded
neither for the knee nor the terminal emission.\\

The axial tail seems to have individual emission `blobs' which appear and disappear or perhaps move. 
An example for the former is the emission in A4 which is prominent in 2007 (epoch B) and in Nov/Dec 2012 (epoch C), but in none of the other epochs. An example for a potentially moving blob is the emission in A2 in epoch D and A3 in epoch E. In Section~\ref{Atail}, we probe the significance of blobs and the hypothesis of outward movement in detail.
In epochs E and F, there is clearly some moderately-hard emission close to the pulsar (in A1). This and the soft A4 emission in epoch C probably cause the observed trend of spectral softening toward the end of the axial tail (Section~\ref{extem}).
The extension of the axial tail never exceeds the length found in the previous observations ($\approx 45\arcsec$), including the merged image. 

\subsection{The immediate surrounding of the pulsar}
\label{marx}
To probe for small-scale structures around Geminga, we simulated the point spread function (PSF) of the pulsar in each observation using MARX\footnote{space.mit.edu/cxc/MARX/index.html} (version  5.0).
There is a known asymmetry region in the {\sl Chandra} PSF\footnote{see cxc.harvard.edu/ciao4.6/caveats/psf\_artifact.html} whose location is different in each of our observations due to changing roll angles.
We only considered individual long ($\gtrsim 60$\,ks) exposures for sufficient count numbers. 
The MARX calibration data is based on CALDB 4.4.7. Since the contamination of the optical-blocking filters of the ACIS detectors\footnote{see e.g., cxc.harvard.edu/ciao/why/acisqecontam.html} is changing the effective area of the instrument over time, mixing of the CALDB versions in the data and MARX simulations would produce inaccurate results. Therefore we extracted the pulsar spectrum for the XSPEC and MARX modeling from data re-processed with CALDB 4.4.7.
We corrected for SIM offsets and used the dither pattern of each observation with an `AspectBlur' of $0\farcs{07}$, which corresponds to the effective blurring of the {\sl Chandra} PSF due to aspect reconstruction
\footnote{cxc.harvard.edu/cal/ASPECT/img\_recon/report.html}
\footnote{space.mit.edu/CXC/MARX/inbrief/news.html}. 
Note that we checked different values of AspectBlur, but obtained only a qualitatively negligible influence on the presence of small structures in the deconvolved images.
We compared total count values of the inner pulsar region of the data and simulation and found a good agreement in the energy range of $0.45$\,keV to 7\,keV. Using the MARX PSF image in this energy range, we deconvolve the pulsar images employing the Lucy-Richardson deconvolution algorithm \citep{Lucy1974}, implemented in the CIAO task \texttt{arestore}. After inspecting different numbers of iterations (10 to 100), we chose $N=30$ as the apparently optimal value. We also checked PSF-subtracted images, but found the deconvolved images to be more useful.\\ 

\begin{figure*}
\begin{center}
\includegraphics[width=170mm]{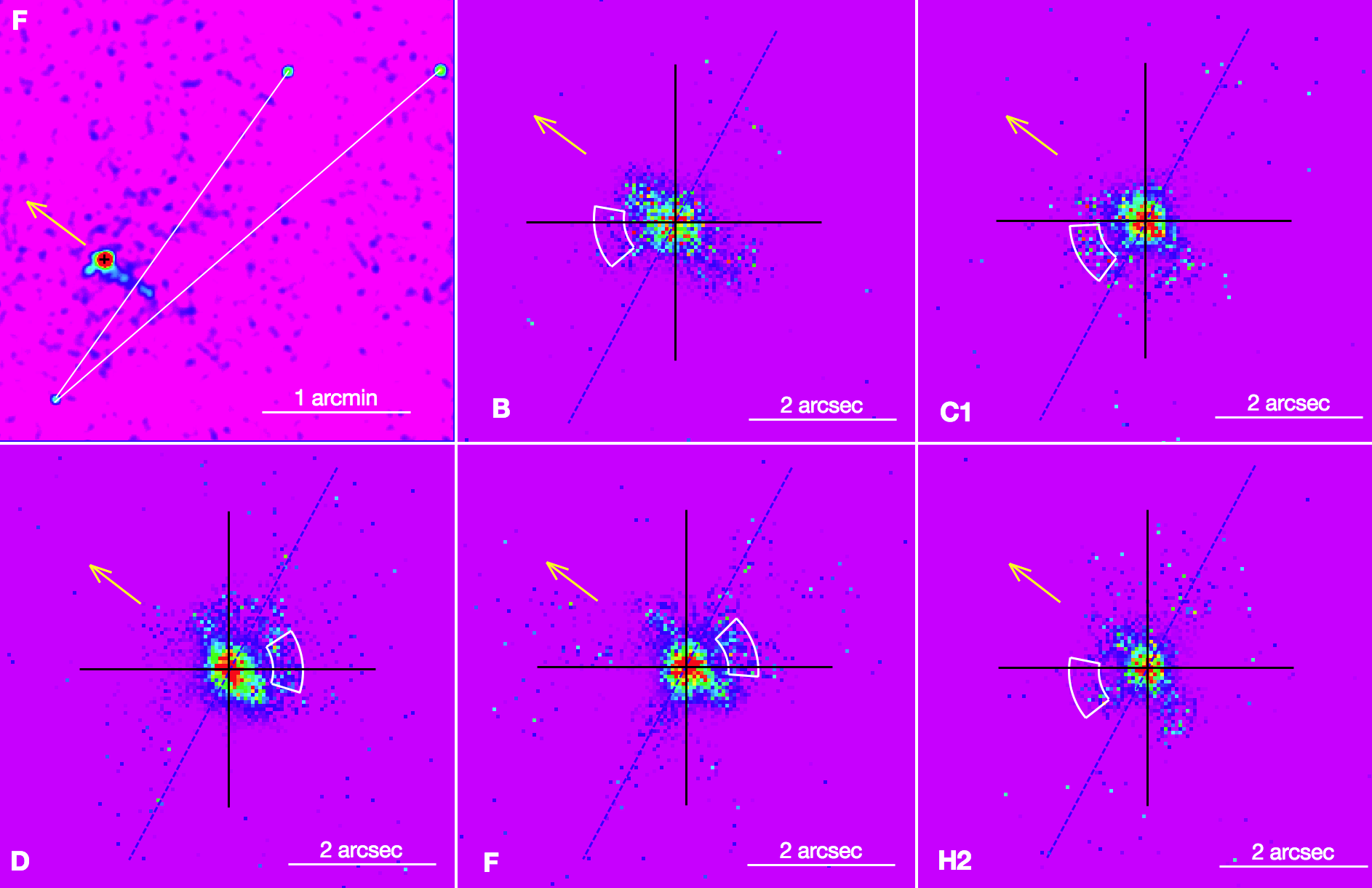}
\vspace{-0.2cm}
\end{center}
\caption{{\sl{Upper left:}} Event file (bin$=1$\,ACIS pixel, smoothed using a gaussian with $\sigma=5$\,pixels) of epoch F showing the location of three reference X-ray point sources with respect to Geminga. These reference stars were used to estimate that the rotational uncertainties are $<1^{\circ}$. {\sl{Zoom-in panels:}} Deconvolved images ($[0.45$\,keV,\,$7$\,keV$]$, bin$=0.1$\,ACIS pixels, not smoothed), obtained with MARX as described in Section~\ref{marx}, are centered on the pulsar in each epoch. North is up, East is to left in all images, these directions are indicated with the big black crosses. The blue dashed line marks the orientation of the footpoints of the lateral tails in epoch F. The white fan-regions show the respective {\sl{Chandra}} PSF asymmetry regions where the count distribution must be regarded with caution.}
\label{marxchanges}
\end{figure*}

Figure~\ref{marxchanges} shows the deconvolved images of the five epochs B, C1, D, F, H2. There are noticeable, varying small-scale structures in these images. Using three reference stars with negligible proper motion in the neighborhood of Geminga (see Figure~\ref{marxchanges}), we confirmed that orientation (rotational) uncertainties are $<1^{\circ}$. We also used the three reference stars to correct for slight shifts.
There are four distinctive emission regions: in front (with respect to the proper motion) of the pulsar, in the direction of the axial tail, and in each direction of the lateral tails N and S. The orientations of these structures appear to change over time. This is most pronounced for the emission in the direction of the S-tail. In epoch F there is a straight south-southeast emission elongation, while in epoch D the S-tail connection is more forward directed (similarly in epoch H2). The angular difference of these orientations is about $40^{\circ}$. Similarly, the connecting emission to the axial tail appears to be more west-directed in epoch B in comparison to the more south-directed emission in epoch H2. The angular difference is at least $\approx 20^{\circ}$. 
It is also noteworthy that the emission in front of the pulsar extends to $\gtrsim 2\arcsec$ in epoch F.\\
 
After our analysis has been completed, a new MARX version appeared (now version 5.3\footnote{space.mit.edu/CXC/MARX/inbrief/news.html\#marx-5-3}) including updates to current CALDB versions as well as important bug fixes regarding the PSF. 
We used one of our longest observations (epoch F, OBSID 14694) to check whether the updated resulting deconvolved image looks different from the one presented in Figure~\ref{marxchanges}. We obtained qualitatively consistent results. In particular, we see the same small-scale extended emission structures in the direction of each tail as well as in front of the pulsar in as seen in Figure~\ref{marxchanges}. Hence, we regard our qualitative findings as robust with respect to the recent CALDB and MARX updates.\\
   
\subsection{The axial tail}
\label{Atail}
While there is a star close to the position of A4, we can exclude that this star is significantly contributing to the axial tail's X-ray emission based on position arguments, the extended nature of A4, and the spectrum and lightcurve of the X-ray emission; for details we refer to the Appendix~\ref{starcontam}.

\subsubsection{Temporal changes of the spatial profile}
\label{axialprofile}
\begin{figure}[h]
\begin{center}
\includegraphics[width=85mm]{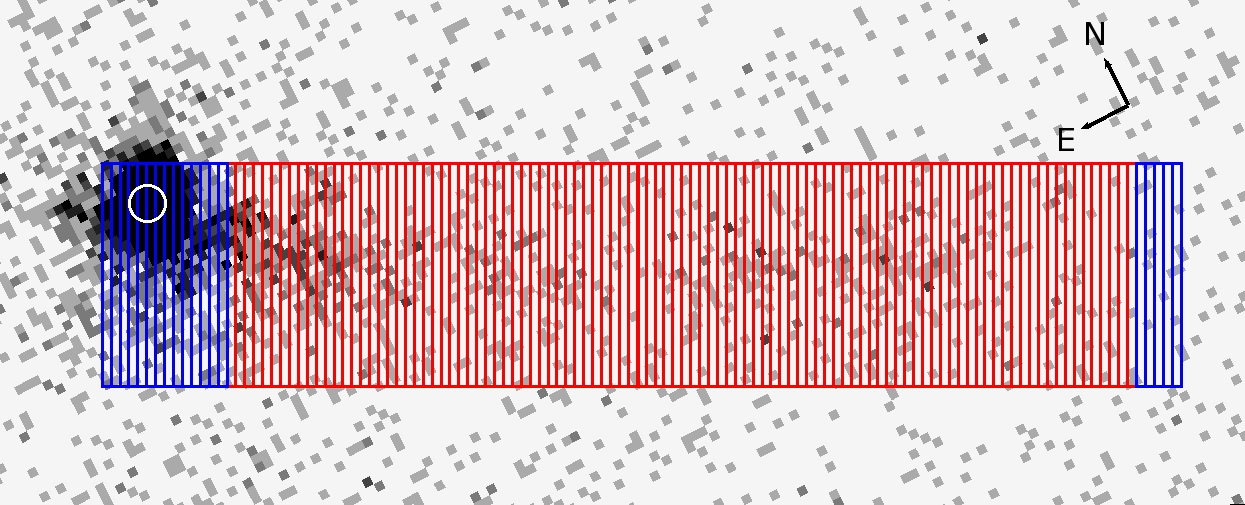}
\end{center}
\caption{Geometry of the analysis region for the linear profile of the axial tail. The background is the merged (epoch C-H) count image. 
Each small count extraction box has a width of 1 ACIS pixel ($0\farcs{49}$) and height of 25 ACIS pixels, the total width of the (blue and red) analysis region is $59\farcs{5}$, the red region, which indicates the plotted range in Figure~\ref{profiles}, has a total width of $50\arcsec$, the circle around the pulsar centroid has a radius of $1\arcsec$.}
\label{boxgeo}
\end{figure}

\begin{figure*}
\includegraphics[width=180mm]{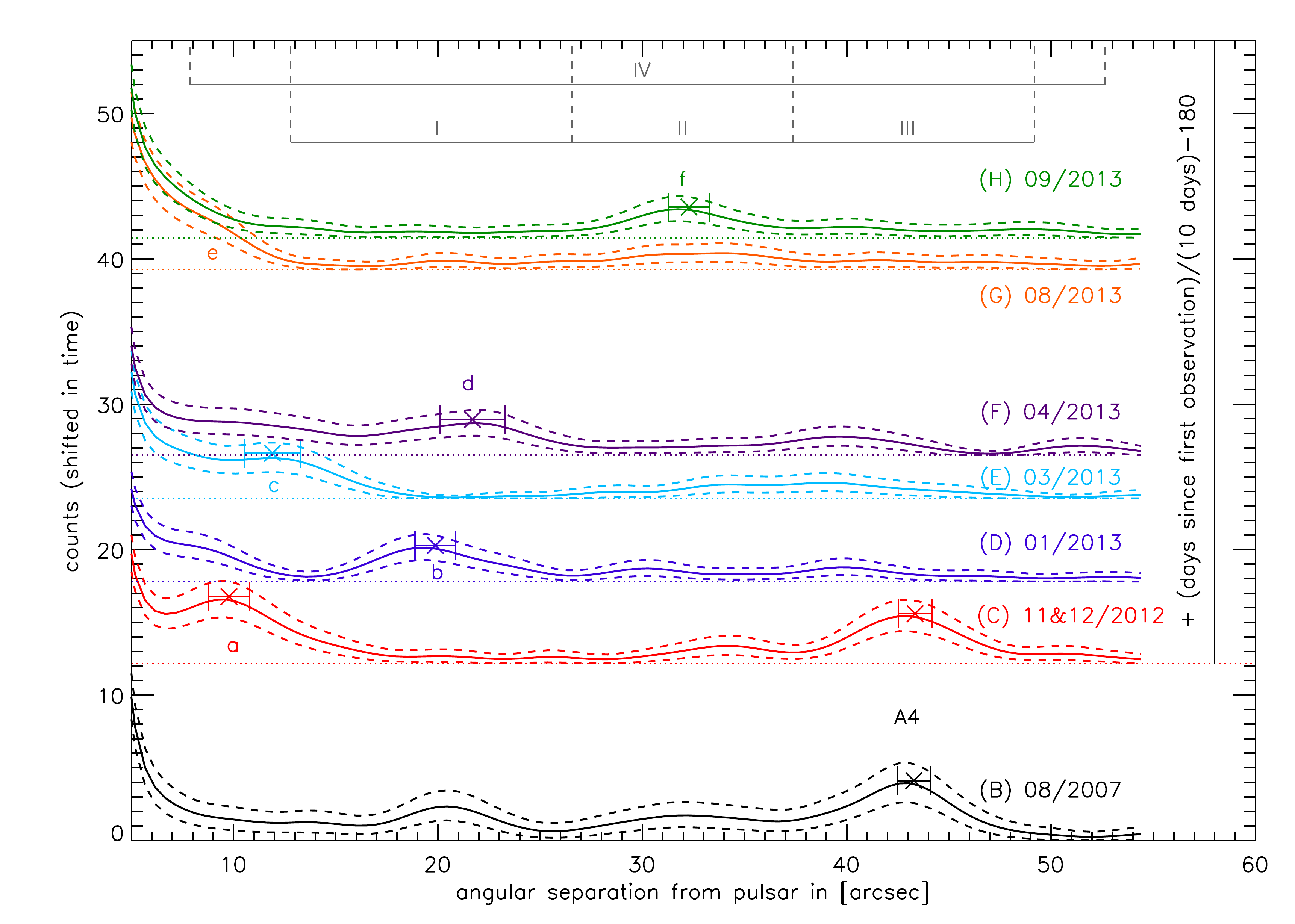}
\caption{Linear count profiles of the axial tail. 
The profiles have been obtained from 10,000 Monte Carlo simulations using the KDE-smoothed original count distribution and total counts in the axial tail in the respective epoch; for details see text.
The solid lines represent the smoothed profiles using a gaussian KDE with width $4\farcs{2}$.
The dashed lines show the 5\% and 95\% quantiles. 
All profiles have been normalized to an exposure time of 100\,ks. Zero levels for the observing epochs in 2012/2013 have been shifted in $y$ by constants proportional to the elapsed time since the 2007 epoch ($\delta y= (t_{i}-t_{2007})/10$\,days$-180$)  and are plotted as dotted lines.  
We also show example $1\sigma$ error bars for blob maxima positions as derived from the Monte Carlo simulations.
Except A4, blob notations do not follow the notations used for the spectral fits in Section~\ref{extem} because A1, A2, A3 are defined on the merged image and might actually consist of overlapping blobs in the individual epochs plotted here.  
The letter 'e' indicates an example of a non-significant peak in the brightness distribution. 
The grey lines and letters on top of the plot show the spatial regions used for the probability analysis in Table~\ref{prob}.}
\label{profiles}
\end{figure*}

Here, we wish to characterise the spatial and temporal evolution of the A-tail.
For each epoch, we extracted the count distribution of the axial tail using thin (1 ACIS pixel or $0\farcs{49}$) boxes along the tail as indicated in Figure~\ref{boxgeo}. 
Note that in the merged count image of Figure~\ref{boxgeo} the A-tail appears to be slightly curved in the vicinity of the pulsar. The count numbers are however too small to measure the curvature.
For our spatial profiles, we adjusted the analysis region locations with respect to the slightly changing pulsar centroid positions at each epoch. 
We used a gaussian kernel density estimator (KDE) for smoothing of the obtained count distribution.
In Figure~\ref{profiles}, we plot the smoothed count profiles of the A-tail for all epochs. 
Note that the actual analysis region is larger (blue in Figure~\ref{boxgeo}) than the range shown for the angular separation from the pulsar in Figure~\ref{profiles} (corresponds to the red region in Figure~\ref{boxgeo}) in order to account for the problematic boundary effects of the KDE.\\ 

To judge the reliability of the obtained profiles and the position uncertainty of blobs (i.e., bumps in the brightness distribution), we carried out bootstrap and Monte Carlo (MC) simulations.
For the bootstrap, we used the maximum entropy bootstrap for time series, the \texttt{meboot}-package in \textsf{R} \citep{Vinod2009,Vinod2013} to produce 10,0000 simulated count distributions for each profile and applied the same gaussian KDE smoothing as previously used for the original data series.
For the MC simulations, we used the original count distribution, smoothed with a gaussian KDE with a width of $3\arcsec$, as probability density function for 10,000 simulations for each observing epoch. The total number of counts measured in the axial tail for a particular epoch defines the sample size in each simulation of that epoch. The same gaussian KDE smoothing was applied to each of the respective 10,000 count distributions as done for the original data series. The MC mean profiles, the 5\% and 95\% quantiles of these 10,000 KDE-smoothed count distributions are plotted in Figure~\ref{profiles} (the bootstrap-profiles are very similar). Note that because KDE smoothing is applied twice during the MC-process, the mean MC-profiles in Figure~\ref{profiles} are the same as if the original series is smoothed once with a gaussian KDE with a width of $3\arcsec \times \sqrt(2) $.
Using the smoothed MC simulations, we also estimated the statistical $1\sigma$ error of selected maxima positions for each epoch.
We found that some smaller blobs in Figure~\ref{profiles} are actually not significant considering count statistics in our simulations. An example is blob `e' close to the pulsar, $\approx 10\arcsec$, in August 2013 (epoch G). While there is clearly enhanced emission in this region, the apparent local maximum of blob e was not distinguishable from  the surroundings when we analysed the 10,0000 MC profiles. Therefore, we concentrate our investigation on few prominent peaks in the profiles as indicated in Figure~\ref{profiles}.\\

Most striking is the presence of strong emission in A4 in 2007 (epoch B) and November-December 2012 (epoch C), but total absence of it at other epochs, in particular at epoch D (see Fig~\ref{profiles}).
Interestingly, the peak position of the central A4 peak in 2007 ($43\farcs{3} \pm 0\farcs{8} $ from the pulsar) is consistent with its position in  2012 ($43\farcs{3} \pm 0\farcs{8} $ from the pulsar).\\

In order to investigate the possibility that some or most of the blobs are just random fluctuations, we use the Anderson-Darling k-sample test, \texttt{ad.test\{kSamples\}} in \textsf{R} \citep{Scholz1987, Knuth2011}.
The Anderson-Darling (AD) statistic tests the hypothesis that several samples all come from the same but unspecified distribution function. 
The AD-test only uses the distribution of the detected count numbers but not their distribution in spatial direction. We split the samples in different subsamples I, II, III, IV as indicated on top of Figure~\ref{profiles} to investigate separate spatial regions. Since the first epoch observation was shorter and has few counts, we checked also whether inclusion of the 2007 data affects the AD statistic or not. The probability results are listed in Table~\ref{prob}. In the most interesting blob regions, I and III, probability values are $\le 6\times 10^{-4}$, proving that the count distributions of the individual epochs do not come from the same parent count distribution.
In particular, blob A4 is not a fluctuation of one parent count distribution function. 
The probability for an origin from the same parent distribution is highest for region II which is not surprising since 'blobs' there are not very pronounced and generally have similarly low count numbers. The probability that the whole (region IV) axial profile (or, more accurately, the count distribution) of all seven epochs are \emph{not} independent from each other is only 1.2\%. Hence, the linear profiles of the Axial tail are indeed inherently different and cannot be explained with random fluctuations.\\  

\begin{deluxetable}{lcc}
\tablecaption{AD-test probabilities\label{prob}}
\tablewidth{8.5cm}
\tablehead{
\colhead{region} & \colhead{6 epochs} &\colhead{7 epochs}}
\startdata
I   & $1\times 10^{-6}$ &  $7\times 10^{-6}$ \\
II  & 0.08 & 0.07 \\
III & $6\times 10^{-4}$ &  $1\times 10^{-5}$ \\
IV  & 0.0016 & 0.0012 
\enddata
\tablecomments{Probabilities for the linear count distribution to come from the same parent count distribution according to the Anderson-Darling k-sample test (see also text). 
The first column lists the spatial regions of the axial tail from Figure~\ref{profiles}. The second and third column list the probability values for the 6-epoch (without 2007)  and 7-epoch (with 2007) samples, respectively. }
\end{deluxetable}

If the axial tail were a jet-like feature, one could expect outward motion of the individual blobs which could be seen in the six most recent epochs. If there were outward motion, one would
expect that the same blob in different epochs would have similar numbers of counts, or maybe slightly fewer in the later epoch because of radiative losses.
Assuming blob `a' ($9\farcs{8} \pm 1\farcs{0}$) in epoch C (red in Figure~\ref{profiles}; live-time-weighted MJD $56260.1$) has moved with a constant velocity to become blob `b' ($19\farcs{9} \pm 1\farcs{0}$) in epoch D (blue in Fig.~\ref{profiles}; MJD $56317$), we estimate an apparent motion of  $0\farcs{18} \pm 0\farcs{02}$\,day$^{-1}$ which corresponds to a tangential velocity of 
$(0.26\pm 0.04) c$. 
Similarly, if we assume that blob `c' ($11\farcs{9} \pm 1\farcs{4}$) in epoch E (light blue in Fig.~\ref{profiles}; live-time-weighted MJD $56374.4$) has moved with a constant velocity to become blob `d' ($21\farcs{7} \pm 1\farcs{6}$) in epoch F (purple in Fig.~\ref{profiles}; MJD $56404$), we estimate an apparent motion of  $0\farcs{33} \pm 0\farcs{07}$\,day$^{-1}$, which corresponds to a tangential velocity of 
$(0.48\pm 0.10) c$. 
These velocities appear to be different, but are still consistent within their $2\sigma$ errors due to the large uncertainties of the peak positions of blobs `c' and `d'.\\ 

We can use the estimated velocities to calculate `expected' previous/future positions of other blobs.
For blob `f' ($32\farcs{3} \pm 1\farcs{0}$) in epoch H (green in Fig.~\ref{profiles}; live-time-weighted MJD $56554.1$) we calculate separations of $28\farcs{4} \pm 1\farcs{1}$, $25\farcs{0} \pm 1\farcs{9}$,  
$\gtrsim 17\farcs{0} $ in epoch G (orange in Fig.~\ref{profiles}; live-time-weighted MJD $56532$) for velocities of $v=(0.26\pm 0.04) c$, $v=(0.48\pm 0.10) c$, $v\lesssim c$, respectively.
Yet there is no prominent blob in epoch G with a similar (or higher) number of counts as detected in blob `f' one month later.\\

For movement of A4 from epoch C, we estimate expected separations of $53\farcs{4} \pm 1\farcs{6}$, $62\farcs{2} \pm 4\farcs{1}$, 
$\lesssim 82\farcs{7} $ 
in epoch D for velocities of $v=(0.26\pm 0.04) c$, $v=(0.48\pm 0.10) c$, $v\lesssim c$, respectively.
Though our profiles in Figure~\ref{profiles} cover only a range up to $55\arcsec$, Figure~\ref{tempchanges} clearly shows that there is no prominent blob within $1.5\arcmin$ in epoch D.\\

In summary, we find indication for different apparent blob velocities, though the significance of the measured difference is low due to small number of blobs we could compare and their low count numbers. 
We conclude that the observed changes in the profile of the axial tail are inconsistent with constant or decelerated motion of emission blobs.
One can, however, also invoke a mechanism for substantial brightening and fading of the blobs on a time scale of a month, see our discussion in Section~\ref{Ataildiscuss}. 

\subsection{The lateral tails}
\label{lattails}
In order to investigate differences between the lateral tails (e.g., their bending) or for a flux comparison between the tails and the cavity between them, it is convenient to use an analytical description for the shape of each lateral tail.
Since the two long tails of Geminga resemble a skewed parabola, we use an analytical expression  $y_0+a (x-x_0)^2$ to fit the tails together and individually (see details in Appendix~\ref{lattailsapp}). Qualitatively, the N-tail has a larger curvature than the  
S-tail: for a fit where the tails connect to the pulsar, $a=0.027$\,arcsec$^{-1}$ for the N-tail, and $a=0.021$\,arcsec$^{-1}$ for the S-tail ($a=0.024$\,arcsec$^{-1}$ if both are fit as one parabola). 

\subsection{The cavity between the lateral tails}
\label{sec:cavity}
\begin{figure}
\includegraphics[height=85mm]{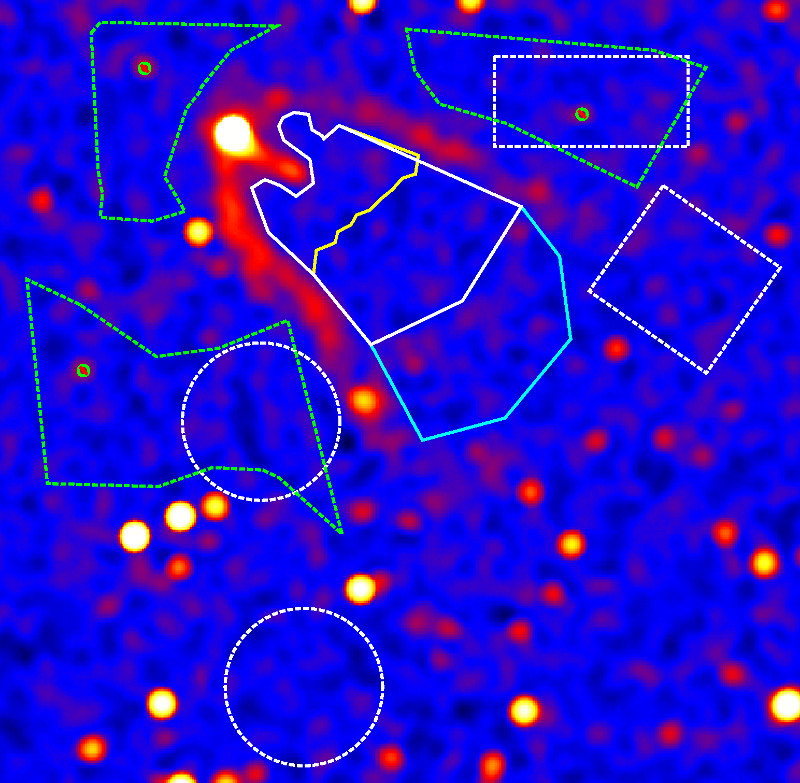}
\caption{Definition of regions used to probe for emission between the lateral tails. 
The yellow small, white medium, and cyan large regions between the lateral tails are investigated for excess emission. 
The green and white dashed regions represent background regions used for the comparison in the merged and the individual observations, respectively (see text).
Regions are overplotted on the same exposure-map corrected flux image as shown in Figure~\ref{Overview}.}
\label{cavity}
\end{figure}

The lateral tails could be limb-brightened outer parts of a shell. 
If so, then we expect excess count numbers over the background between the tails due to emission projected from the front and back of the shell. 
In order to probe this hypothesis, we investigated the exposure-corrected {\emph{merged}} (2012-2013) image in the energy range $0.5-7$\,keV. In addition, aiming to minimize systematic effects due to the different localization of the target on the ACIS chips, we investigate the exposure-map-corrected images of 8 observations with comparable localizations.
For the merged data, we compared the average flux values per pixel, $\bar{F}_{-11}$, in units of $10^{-11}$\,photon\,cm$^{-2}$\,s$^{-1}$\,pix$^{-1}$ (a pixel in this case is $0\farcs{984}\times 0\farcs{984}=4$\,ACIS pixels), of background regions and three `cavity' regions of different size (see Figure~\ref{cavity}). Since one deals with small count numbers and hence strong Poisson noise, we used the actual total count numbers, $N_C$, in the same regions to derive (minimum) error estimates of the average flux values, $\delta \bar{F}_{-11}= \bar{F}_{-11} \times N_C^{-1/2}$.
From the three background regions (green, Figure~\ref{cavity}) we obtained  $\bar{F}^{\rm Bgr}_{-11}=  230 \pm 2$. For the small (yellow), medium (white), and large (cyan) `cavity' regions we derived $\bar{F}^{\rm Cav}_{-11}= 237 \pm 6$, $233 \pm 4$, and $236 \pm 3$, respectively. Thus, we measure a maximal significance of $1.7\sigma$ for any flux enhancement for the merged data set.\\

Employing different background regions for one of the long (100\,ks) exposures, we noticed some scatter between average flux values (e.g., the significance of the difference reached $2.2\sigma$) at different background locations. CCD gaps in particular seemed to cause deviating values. The merged Geminga data set was derived from observations where the CCD gaps are located at different sky positions. To minimize systematic effects due to placing of the extended emission (and background), we used 8 observations with similar setup, namely epochs B, C1, C2, G1, G2, G3, and H2.
For each observation, we followed the same procedure as for the merged data set. We used, however, the white background regions and the white medium-sized `cavity' region in Figure~\ref{cavity} in order to avoid CCD gaps.
The epoch-B--values strongly deviated from the other epochs and were excluded.
We derived the live-time--weighted mean, the median, and the standard deviation of the remaining 7 average flux values. For the background, we obtained  $\bar{F}^{\rm Bgr}_{-11} (\rm weighted) =  228$,  ${\rm median}({F}^{\rm Bgr}_{-11}) =  226$, ${\rm stddev} {F}^{\rm Bgr}_{-11} =  9$. For the medium `cavity' region, we obtained $\bar{F}^{\rm Cav}_{-11} (\rm weighted) =  221$,  ${\rm median}({F}^{\rm Cav}_{-11}) = 221 $, ${\rm stddev} {F}^{\rm Cav}_{-11} =  11$.
Using this more conservative method, we do not detect any flux enhancement at the `cavity' region. Based on this result and the low-significance result for the merged data, we conclude that there is no detectable flux enhancement in the investigated `cavity' region.\\ 

\begin{deluxetable*}{lccccccccccc}[h!]
\tablecaption{Magnetic field and flow parameters assuming equipartition between particle and magnetic energy\label{Bestimates}}
\tablewidth{0pt}
\tablehead{
\colhead{Region} & \colhead{$\Gamma$} & \colhead{$p$} & \colhead{$b_{\rm min,p}$} & \colhead{$b_{\rm max,p}$} & \colhead{$a_p$} & \colhead{$\bar{s}_{16}$} & \colhead{$\mathcal{B}_{-7}$} & \colhead{${B}$} & \colhead{$\gamma_{\rm char}$(6\,keV)}  & \colhead{$l_{17}$}& \colhead{$v_{\rm flow}$}} 
\startdata
S-tail & 1.04 & 1.08 & 0.80 & 0.00045 & 0.248 & 8.00 &  0.00692 & $20$\,$\mu$G & $2.4\times 10^8$ & 6.7 & $>1260$\,km\,s$^{-1}$\\
A-tail & 1.63 & 2.26 & $\approx 2.0$ & $\approx 0.066$ & 0.091 & 3.74 &  0.05242 &  $21$\,$\mu$G  & $2.3\times 10^8$ & 1.6& $>340$\,km\,s$^{-1}$\\
\enddata
\tablecomments{see text for explanation of quantities and units}
\end{deluxetable*}

In order to roughly estimate the lower limit of the observed ratio of the lateral tails to the cavity, we used the same procedure for the N- and S-tails for the merged data set and derived
$\bar{F}^{\rm N}_{-11}= 348 \pm 8$ and $\bar{F}^{\rm S}_{-11}= 413 \pm 8$. Considering $\bar{F}^{\rm Bgr}_{-11}=  230 \pm 2$ and a $3\sigma$ upper limit for the detection of the (medium-sized) `cavity', 
we estimated $\bar{F}_{\rm bgr.cor.}^{\rm N}/\bar{F}_{\rm bgr.cor.}^{\rm Cav}> 7.4$ and  $\bar{F}_{\rm bgr.cor.}^{\rm S}/\bar{F}_{\rm bgr.cor.}^{\rm Cav}> 11.6$, and $>9.1$ if both tails are considered together.\\

For a simplified model of a shell with locally isotropic (and optically thin) emission, one can estimate the \emph{expected} emission ratio between the lateral tails and the cavity by using the corresponding volume ratio of an approximate parabolic shell whose envelope is the shape spanned by the lateral tails (Section~\ref{lattails}). The thickness, $s$, of the parabolic shell is approximated as the observed average thickness of the lateral tails, 
$21.4\arcsec$ ($0.026$\, $d_{250}$\,pc).
In case of synchrotron emission, of course, the assumption of locally isotropic emission is only valid if the orientation of the magnetic field is random in the emitting region. This might or might not be the case, and we emphasize this limitation of this toy model.
Cutting the apex from the top of the parabola, we estimate the shell volume that gives the lateral tails, $V_{\text{lat.tails}}$, and the shell volume that contributes to the emission in between the tails (named above the `cavity'), $V_{\text{cav}}$. Employing the analytical desciption of the parabola (see Appendix~\ref{lattailsapp}), we use different integration intervals along the symmetry axis for $V_{\text{lat.tails}}$ and $V_{\text{cav}}$. This allows us to compare the analytical result with the measured flux for which we have to avoid the axial tail. 
The expected value of this volume ratio 
is
\be
\frac{V_{\text{lat.tails}}(18\arcsec,175\arcsec)}{V_{\rm cav}(49\arcsec,158\arcsec)}=0.9.
\label{calcresult}
\ee
Using the \emph{observed} fluxes and uncertainties in the corresponding regions, and 
employing the same method as above to estimate the $3\sigma$ upper limit, we obtain 
\be
\frac{F_{\rm BGcor}^{\rm lat.tails}(18\arcsec,175\arcsec)}{F_{\rm BGcor}^{\rm Cav}(49\arcsec,158\arcsec)} > 4.5.
\label{mcavres1}
\ee
Thus, the observed flux ratio between lateral tails and cavity is at least a factor 5 larger than the flux ratio expected from locally isotropic emission in a parabolic shell with thickness of
$21.4\arcsec$ ($0.026$\, $d_{250}$\,pc).

\section{Discussion}
PBZ10 discussed in detail possible explanations for the observed morphology of the Geminga PWN. In our discussion we focus on the additional insights gained by our new observations.
The new contraints on the physical length and spectra of the extended emission structures, for example, allow us to estimate the magnetic field for each tail. For this, we assume synchrotron emission from particles with a power law energy distribution with index $p$ which is related to the measured photon index $\Gamma$ by $p=2 \Gamma - 1$. The magnetic field $B$ is estimated according to basic equations for synchrotron radiation \citep{Ginzburg1965} as: 
\be
\label{magformula}
B=27 \left( \frac{k_m}{a_p (3-2\Gamma)} \left[ E^{(3-2\Gamma)/2}_{\rm max,p} - E^{(3-2\Gamma)/2}_{\rm min,p}  \right] \frac{\mathcal{B}_{-7}}{\bar{s}_{16}}\right)^{2/7} \mu \mathrm{G}
\ee
where $k_m=w_{\rm mag}/w_{\rm rel}$ is the ratio of the magnetic energy density to the energy density of the relativistic particles, i.e., $k_m \approx 1$ if equipartition is assumed. 
$E_{\rm max,p}=E_{\rm max}/b_{\rm max,p}$ and  $E_{\rm min,p}=E_{\rm min}/b_{\rm min,p}$ are related to the chosen energy bounds $E_{\rm max}=8$\,keV and $E_{\rm min}=0.3$\,keV, while $b_{\rm max,p}$, $b_{\rm min,p}$, and $a_p$ are numerical coefficients from \citet{Ginzburg1965}. 
The average spectral surface brightness at energy $E=1$\,keV is
$\mathcal{B}=\mathrm{PL Norm}/\mathrm{Area}=10^{-7}\mathcal{B}_{-7}$ photons (s\,cm$^2$\,keV\,arcsec$^2$)$^{-1}$. 
The average length of the radiating region along the line of sight is
$\bar{s}=10^{16} \bar{s}_{16}$\,cm, 
and we approximate $\bar{s}$ with the average observed ``thickness'' of the tails assuming cylindrical symmetry ($\approx 21\arcsec$ for the S-tail, $\approx 10\arcsec$ for the A-tail). 
In Table \ref{Bestimates}, the values of the respective quantities are given for the A-tail and S-tail\footnote{{The formula cannot be applied to the N-tail since 
its photon index, $\Gamma = 0.67\pm 0.12$,
is close to the minimum possible value, $\Gamma_{\rm min}=2/3$, for 
optically thin synchrotron radiation for any electron distribution
(Ginzburg \& Syrovatskii 1965).}}. 
The inferred magnetic field strengths of S- and A-tail are comparable to those of other PWNe if equipartition is assumed there (e.g., \citealt{Auchettl2015,Reynolds2012,Pavlov2003}).
Assuming synchrotron cooling, we can only estimate lower limits on the velocity in the bulk flows of the tails, $v_{\rm flow}$, because we do not see any significant spectral softening for any of the tails. 
For the tail lengths, $l=10^{17} l_{17}$\,cm, and characteristic Lorentz factors, $\gamma_{\rm char} (E) \sim 1.4 \times 10^8 (B/10\mu G)^{-1/2}(E/{\rm 1keV})^{1/2}$ at $E=6$\,keV, the derived limits of the (projected) flow velocity are $v_{\rm flow}>1260 d_{250}$\,km\,s$^{-1}$ for the S-tail, and $v_{\rm flow}>340 d_{250}$\,km\,s$^{-1}$ for the A-tail.
For comparison, the speed of knots in the Vela jet have been estimated to be $0.3 c-0.7c$ \citep{Pavlov2003}.

\subsection{Possible Interpretation of the Lateral Tails}
The outer tails could represent either a limb-brightened shell or bent collimated outflows.
The new data allow us to better constrain the shape formed by the lateral tails (Section~\ref{lattails}), potential emission in the cavity between the tails (Section~\ref{sec:cavity}), the spectra of the lateral tails (Section~\ref{extem}), their changes over time (Section~\ref{sec:tempchanges}) and their connection to the pulsar (Section~\ref{marx}). The patchiness of the lateral tails, already observed by PBZ10, is confirmed by the new data (see Figure~\ref{tempchanges}). However, count numbers in individual ``patches'' of the lateral tails are still too small to detect potential motion of these patches at a useful statistical significance level, in particular taking background fluctuations into account.\\  

A shell could be formed by ram pressure of the ISM impinging on the pulsar wind which could be  isotropic or equatorially concentrated.
If the lateral tails represent the limb-brightened boundaries of such a projected shell, PBZ10 concluded that the emission might be due to synchrotron radiation from the region where shocked pulsar wind (PW) and shocked interstellar medium (ISM) mix. There, the PW is decelerated to non-relativistic bulk velocities by mass loading due to the shear instability. 
PBZ10 suggested that the shape of the shell is different from the surface of the contact discontinuity (CD) in available numerical PWN models because these models do not include mass loading and the proper anisotropy of the unshocked PW.\\ 

According to a recent paper by \citet{Morlino2015}, mass loading in PW tails with \emph{neutral} hydrogen with a density as low as $10^{-4}$\,cm$^{-3}$ is expected to produce a secondary shock with resulting fan-shaped 
Balmer emission opposite to the direction of pulsar proper motion (``behind the pulsar''). 
For Geminga, H${\alpha}$ emission behind the pulsar was seen neither by \citet{Caraveo2003} in their VLT FORS1 image, nor in the similarly deep IPHAS survey \citep{Drew2005,Barentsen2014}. 
According to \citet{Giacani2005}, Geminga is apparently located in a local minimum of neutral hydrogen. This explains a missing H${\alpha}$ PWN for Geminga and calls into question a neutral hydrogen mass loading scenario.\\

Any interpretation of the lateral tails should take into account their 
unusually hard spectra (Table~\ref{extfit}, Figure~\ref{contours}).
Their low $\Gamma \lesssim 1$ is clearly different from those of the axial tail or the
PL-component of the pulsar's spectrum.
Considering the jet interpretation of the lateral tails, the hard emission is not unprecedented 
(e.g., the Vela jet has $\Gamma=1.3\pm 0.1$, \citealt{Pavlov2003}). However, such hard emission seems unusual in common bow shock models, hence in the case of the shell interpretation.  
However, the very hard spectral index of the N-tail ($\Gamma <1$) already indicates that the spectral energy distribution of the X-ray emitting electrons is possibly not a PL and/or may be produced by a supplemental acceleration mechanism.
If one assumes that the hard emission is of synchrotron origin, a very hard PL
spectral energy distribution (SED) would be required, $dN_e/d\gamma \propto \gamma^{-p}$ with $p\approx 1$. The commonly considered Fermi acceleration mechanism at fronts of relativistic shocks gives $p\gtrsim 2$, which corresponds to 
$\Gamma\gtrsim 1.5$ (see, e.g., Chapter 6 of the review by \citealt{Bykov2012}). 
However, a harder SED with $p\approx 1$ can be produced 
by the Fermi mechanism at the shocks that form in two
colliding MHD flows.
\citet{Bykov2013} simulated particle acceleration in such a system 
for a nonlinear model that includes the back-reaction 
of the accelerated particle pressure, in a simplified one-dimensional geometry.
A very hard spectrum of accelerated electrons with a slope $p\approx 1$
has been revealed at the high-energy end of the SED.
In the case of a supersonically moving pulsar, such as Geminga, the two colliding
flows (in the reference frame of CD) are the relativistic
PW and the oncoming ambient medium with density ${\rho}_{\rm a}\approx n_{\rm a} m_{\rm P}$ ($m_{\rm P}$ is the mass of a proton).
The hard spectrum with $\Gamma\approx 1$ could be produced in the region
between the PW termination shock and the
forward bow shock (i.e., around the CD), and has a characteristic size comparable to 
the stand-off radius 
\be
\label{standoff}
R_s=\left[ \dot{E}_{\rm PW} f\, {(4 \pi c \, {\rho}_{\rm a} \, v^2_{\rm total})^{-1}} \right]^{1/2},
\ee
where $v_{\rm total}$ is the total velocity of the pulsar, 
$\dot{E}_{\rm PW} = \xi_w \dot{E} \lesssim \dot{E}$
is the spindown power emitted with the wind, and the factor $f$ takes into account possible anisotropy of the pulsar wind ($f=1$ for an isotropic wind).
To be accelerated by the Fermi mechanism between the converging flows, there
should be high-energy electrons in the PW that have a mean free path exceeding the distance between the two shocks.
Employing the gyroradius of the electrons, $r_g(\gamma) = \gamma m c^2/eB$, the maximal energies of these accelerated electron are limited by the condition
$r_g(\gamma_{\rm max}) \lesssim R_{\rm s}$. 
For the parameters of Geminga, 
$\gamma_{\rm max} \sim 6\times 10^7 (\xi_w f)^{1/2} n_{\rm a}^{-1/2} (v_{\rm PSR}/211\,{\rm km\,s}^{-1})^{-1} B_{-5}$ corresponds to synchrotron photon energies $E\sim 0.2 \xi_wf B_{-5}^3 n^{-1}_{\rm a}$ keV.
Thus, to have the maximum photon of about 10 keV (the upper energy limit of the \emph{Chandra} observations), we require $\xi_w f B_{-5}^3 \gtrsim 50 n_{\rm a}$, i.e., sufficiently low ambient densities.
This result demonstrates that the Fermi acceleration at the shocks of two colliding MHD flows can indeed operate for Geminga in the limb-brightened shell interpretation.\\ 

Although the hard X-ray spectrum of the lateral tails could be explained
in the limb-brightened shell interpretation by the
Fermi acceleration at the two colliding shocks, the nondetection of 
shell emission between the tails makes this interpretation questionable if the shell's shape is close to a body of rotation with symmetry axis along the pulsar's proper motion direction (e.g., paraboloid).
The new observations showed that the ratio of surface brightness (or fluxes) between the outer tails and the cavity is at least a factor 7 and 12 for the N- and S-tail, respectively (Section~\ref{sec:cavity}). 
Such ratios are difficult to reconcile with the shell interpretation if the magnetic field  is randomly oriented in the emission region. 
However, if the strength of magnetic field 
varies as a function of the azimuthal angle around the shell axis, then 
different parabolic regions along the shell surface may have different 
brightness (as the synchrotron emissivity is proportional to $B^{p+1} \approx B^2$. Such an azimuthal dependence may be caused by an amplification of the ISM magnetic field component parallel to the forward shock surface, which could be up to a factor of 4 for a large Mach number.
For instance, if the ISM field is directed along the line of sight and the pulsar velocity is
nearly perpendicular to the line of sight, the \emph{amplified} magnetic field is stronger at 
the shell limb, which leads to an enhanced shell brightening. One would still expect the `head' of the shell to be filled with X-ray emission though. From Figure~\ref{cavity}, this does not appear to be the case, but we cannot reliably assess the emission between the lateral tails close to the pulsar because of low count numbers and the presence of the axial tail.\\

Another explanation for the lack of detectable emission between the lateral
tails in the shell interpretation could be a strong azimuthal asymmetry of the
PW with respect to the pulsar's direction of motion. Indeed, if the PW is
concentrated in the equatorial plane and the pulsar's spin axis is
misaligned with the direction of motion, then the shell would not be a body
of rotation. In the extreme case when the spin axis is perpendicular to the
direction of motion and parallel to the line of sight (i.e., the
equatorial plane is in the plane of the sky), one would not expect to see a
limb-brightened axially-symmetric shell. Instead, two bent streams with no
(or little) emission between them, could be an observational possibility which needs to be further investigated with detailed simulations that are beyond the scope of this paper.\\  

In the standard bow shock picture, one would expect to see the shell apex at some distance ahead of the moving pulsar (``in front of the pulsar''), however the lateral tails not only seem to connect directly to the pulsar but also the emission of their footpoints seem to change orientation (Figures~\ref{tempchanges} and \ref{marxchanges}).
This is most prominent when comparing epochs D and F, e.g., in Figure~\ref{marxchanges}. We emphasize that these two epochs are the longest single exposures available, representing the best count statistics on the footpoints of the lateral tails.
X-ray emission in front of the pulsar in the Ring ($3\arcsec - 8\arcsec$) and Bow regions ($8\arcsec - 16\arcsec$) -- Figure~\ref{Overview} -- is consistent with the spatial count distribution tail from the point source for on-axis imaging with the HRMA/ACIS.
However, the investigation of the immediate surrounding of the pulsar revealed several indications of close emission in front of the pulsar, e.g., epochs B and F in Figure~\ref{marxchanges}, E and H in  Figure~\ref{tempchanges}. This emission appears to be oriented slightly differently in each epoch. If this varying emission is due to fluctuations (either count fluctuations or actual physical variations) of the CD surface head, the latter would be very close ($\lesssim 1\arcsec$) to the pulsar. 
In the framework of \emph{isotropic} PWN models, a very close CD surface head is difficult to reconcile with the large spatial separation of the two lateral tails, but it is possible for an equatorially confined pulsar wind if the equatorial plane is nearly perpendicular to the proper motion direction (e.g., \citealt{Bucciantini2005, Vigelius2007, Romani2010}).\\

On the other hand, we may interpret the lateral tails as collimated outflows,
i.e. polar jets along the pulsar's spin axis. Such jets are seen in many PWNe (e.g. \citealt{Kargaltsev2008});
they appear to suffer deflection by both internal instabilities and sweepback by external
ram pressure. In Geminga, the latter clearly dominates, and we expect the entire
shocked pulsar wind to sweep back along the line of motion, inside the CD. As polar jets, the lateral tails exhibit continued collimation
and
continuous re-injection of fresh energetic particles which may explain the hardness of their spectra.
In this view these bright structures
should follow the downstream flow of the shocked pulsar wind, remaining within
the CD. In some objects where we can see the forward ISM shock
in H$\alpha$, e.g. PSR J2124$-$3358, the jet remains embedded in the shocked pulsar wind. A full MHD model with jets misaligned to the
pulsar motion would be required to study this behavior.
An alternative picture arises if the jets represent \emph{highly collimated} momentum flux, similar to the outer Vela jet (whose formation mechanism is not understood yet). In this case the jets may propagate through the pulsar wind, CD
and forward shock and then suffer gradual sweepback as they encounter the
ram pressure of the general ISM. This is similar to the picture for AGN jets embedded in
cluster gas (e.g., \citealt{Odea1985}).
As outlined in Appendix \ref{jetsapp}, the momentum flux imparted by the ISM in the neighborhood of Geminga is sufficient to
sweep back such highly collimated polar jets.\\ 

As discussed by PBZ10, the interpretation of the lateral tails as bent jets would imply a large angle $\theta$ between the spin axis and the direction of the pulsar velocity; based on Figure~\ref{marxchanges}, $\theta$ is between $45^\circ$ and $80^\circ$. 
The angle $\zeta$ between the direction of sight and the spin axis also needs to be sufficiently large to explain a similar bending of the two jets. A large $\zeta$ would agree with geometric constraints from Geminga's pulses (see Section~\ref{intro}). The N-tail appears to be slightly more bent than the S-tail, though (see also profile fits listed in Table~\ref{lattailfit}). This can be partly due to geometric projection. The lateral tails can also trace inhomogeneities in the ISM (in the jet as well as shell interpretation).   
As PBZ10 already noted, Geminga is surrounded by an incomplete ring of H{\sc{i}} emission (with average radius of $9\arcmin$) which is open in the northwest \citep{Giacani2005}. The bright S-tail coincides with the border of the H{\sc{i}} shell.\\

The N-tail is fainter than the S-tail. 
The brightness difference might be caused by Doppler boosting unless $v_j \ll c$.  
Since the alleged jets are strongly bent, the angle between the bulk motion velocity component and the line of sight change over the length of each jet. One would expect brightness differences due to Doppler boosting along the length of each jet, too. 
Yet, there are no prominent brightness changes along the lateral tails in the {{merged}} data, in particular the brighter S-tail, see, e.g., first panel in Figure~\ref{tempchanges}. 
This indicates that the alleged jets are located in a plane almost perpendicular to the line of sight, hence $\zeta \approx 90^\circ$.  
Precessing jets could introduce additional time-variable changes of brightness pattern due to Doppler boosting.
The patchy pattern seen in the lateral tails in individual epochs in Figure~\ref{tempchanges} do not have enough counts to probe such a hypothesis.
Another explanation for the brightness difference between N- and S-tail could be intrinsically different jets which could be caused, for example, by a de-centered magnetic dipole field.\\

If the lateral tails indeed represent polar outflows, then the implied axes orientation would place Geminga in the tail of the statistical $\theta$ distribution by 
\citet{Noutsos2012} who found  strong evidence for a general alignment of a pulsar's spin axis with its velocity axis from a study of 54 pulsars. 
While such an axis orientation is unusual, in particular for ``young'' pulsars ($< 10$\,Myr; \citealt{Noutsos2013}), it is possible to explain. For example, \citet{Kuranov2009} reported that tight spin-velocity alignments are more probable for single progenitors while binary progenitors will more likely result in larger $\theta$.\\ 

\subsection{Possible Interpretation of the Axial Tail}
\label{Ataildiscuss}
As discussed by PBZ10, the axial tail could be a pulsar jet
or a shocked PW. The new data showed individual, short-lived ($\lesssim 1$\,month) blobs in the axial tail and enable their spectral analysis (albeit hampered by few counts; Section~\ref{extem}), and an investigation for temporal changes 
(Section~\ref{axialprofile}). Deconvolution with the MARX-simulated {\sl Chandra} PSF also provided a view of the axial tail in the immediate pulsar vicinity as well as of emission in front of the pulsar  (Figure~\ref{marxchanges}).\\

If the axial tail is a jet,
one would expect the blobs of the axial tail to move with mildly relativistic velocity. In Section~\ref{axialprofile}, it is shown that the blobs appear to have different velocities if outward motion is assumed. Moreover, blobs seem to brighten or get fainter on a time scale of a month. Based on our analysis in Section~\ref{axialprofile}, we conclude that the temporal changes of the blobs in the axial tail do not support an interpretation of constant or decelerated motion away from the pulsar. 
A jet
appears therefore an unlikely explanation for the axial tail, but cannot entirely be excluded due to possible perturbations in the flow which could destroy blobs on time scales $\gtrsim 10$\,days.\\ 

Recent MHD simulations for the Crab PWN by \citet{Porth2014} confirmed the importance of the anisotropic structure of pulsar winds to reproduce the PWN's torus and jet structures. 
For the middle-aged, fast moving Geminga, such anisotropic PW could in principle produce a distorted (or even crushed) PW torus from an equatorially confined wind, which could explain the axial tail and the emission in front of the pulsar. In such interpretation the lateral tails would be polar outflows. According to equation~\ref{standoff} one would expect bright emission from the distorted torus shock in front of the pulsar at a standoff distance $R_s= 1.1\times 10^{16} (f \xi_{\rm PW})^{1/2} n^{-1/2}_a d^{-1}_{250}$\,cm. 
From jet bending, we estimated $n_a< 0.007 \text{cm}^{-3}$ (for a highly collimated jet with $\xi_j=0.1$, see Appendix~\ref{jetsapp}), hence $R_s> 34\arcsec (f \xi_{\rm PW})^{1/2}$.
However, no bright emission is observed at such large separation in front of the pulsar (see also previous section). Instead, Figure~\ref{marxchanges} shows that there is emission very close ($<2\arcsec$) in front of the pulsar. This emission seems to change direction (comparing, e.g., epochs C1 and F), and to have sometimes larger extensions (e.g., epoch H, and epochs G and E in Figure~\ref{tempchanges}). While a low $\xi_{\rm PW}$ will lead to a lower $R_s$, it remains difficult to explain why there is no bright X-ray emission at larger separations ahead of Geminga. In this respect, it is also interesting to note that luminosities of the tails, $\approx 1.6 \times 10^{29}$\,erg\,s$^{-1}$, $\approx 2.6 \times 10^{29}$\,erg\,s$^{-1}$, $\approx 0.9 \times 10^{29}$\,erg\,s$^{-1}$ for the N-, S-, and A-tail respectively, indicate a rather high luminosity ratio ($\approx 5$) of jet to torus. Usually, PW torii are much brighter than the pulsar jets (e.g., Vela).\\ 

If the axial tail is interpreted as the CD-confined cylindrical region behind an unresolved termination shock of an equatorially confined wind, the blobs could be due to, e.g., shear instabilities at the CD surface. However, it seems a strange coincidence that blob A4 is prominent at a similar position with a similar brightness in 2007 and Nov/Dec 2012, but disappeared completely in January 2013 and all other epochs. 
Another possible explanation for the blobs could be plasmoids formed by magnetic reconnection. 
Such structures are known for the magnetotail of the Earth (for recent review see, e.g., \citealt{Eastwood2015,Eastwood2015b}). 
In the case of the Earth, magnetic reconnection across the magnetotail's current
sheet create a changing pattern of magnetospheric convection zones, a process known as the Dungey cycle \citep{Dungey1961}. 
A pulsar with its wind, moving through the ISM, shows some resemblance to the Earth's magnetosphere encoutering the solar wind.
Recently, \citet{Sironi2016} carried out large-scale two-dimensional particle-in-cell simulations in electron-positron plasmas and demonstrated that relativistic magnetic reconnection can also lead to the formation of quasi-spherical plasmoids filled with high-energy particles and magnetic fields.
As outlined by \citet{Sironi2016}, future studies of such plasmoids in 3D with consideration of radiative cooling are needed for actual quantitative constraints of the plasmoid properties in PWNs and relativistc jets.\\

\section{Conclusions}
The six new {\sl Chandra} observation epochs of the PWN around Geminga have resulted in the following firm observational findings:
{\emph{(i)}} The overall morphology of Geminga's PWN does not change with the six times deeper image. There are two $\sim 3\arcmin$ long lateral tails and a segmented axial tail of about $45\arcsec$ length. 
{\emph{(ii)} There is no detected X-ray emission between the lateral tails. The
ratio of surface brightness between the outer tails and the cavity is at least a factor 7 and 12 for the N- and S-tail, respectively.
{\emph{(iii)} The axial tail consists of individual emission blobs at different separations from the pulsar. These blobs appear and disappear on time scales of a month. There is no convincing evidence for constant or decelerated movement of these blobs. 
{\emph{(iv)} The lateral N-tail shows a stronger bending than the S-tail.
{\emph{(v)} The lateral tails have significantly harder spectra than the axial tail or the magnetospheric emission of Geminga itself.\\

Less firm, due to potential unknown systematics in the image deconvolution analysis are the following findings:
{\emph{(vi)}} The lateral tails seem to directly connect to the pulsar. Their footpoints seem to ``wiggle'' when comparing individual epochs. 
{\emph{(vii)}} There is no bright arc-like emission feature in front of the pulsar.
{\emph{(viii)}} There is, however, protruding X-ray emission very close ($<2\arcsec$) in front of and also behind the pulsar. This emission is differently pronounced in the individual epochs and possibly wiggles too.\\

Several physical models are still possible for the interpretation of the Geminga PWN. 
The shell interpretation for the lateral tails (and a jet-like outflow confined by the ISM ram pressure for the axial tail) requires either an ISM magnetic field oriented perpendicular to the line of sight and amplification of the magnetic field or an azimuthally asymmetric shell. The explanation for the hard emission of the lateral tails within this model also leads to the question why the Fermi acceleration mechanism in colliding winds does not produce similarly hard emission in other pulsar bow shocks.
The jet interpretation for the lateral tails (and equatorial outflow for the axial tail) would require unusually luminous jets. Within this interpretation, it remains puzzling why no prominent emission is observed ahead of the pulsar at separations $>2\arcsec$.
Currently, it is not possible to rule out either of these two scenarios for the lateral tails.\\

In order to ultimately understand the physics of the enigmatic Geminga PWN, MHD simulations of a fast-moving pulsar with an anistropic PW would be extremely helpful. 
Observationally, new insights into the Geminga PWN could be gained by X-ray polarimetry observations targeting the magnetic field orientation which governs the PWN shape. It would be also interesting to know the magnetic field orientation in the ISM around Geminga. In principle, such knowledge could come from refined local maps of the polarized thermal emission from Galactic dust \citep{Planck2015xix} or results from on-going and planned Galactic radio polarimetry surveys as outlined by \citet{Haverkorn2015}.
Future high resolution, high statistics X-ray observations may
be able to directly probe the fine structure and instabilities of this fascinating nearby PWN, in particular in the region close to the pulsar. However, such observations will likely require substantial increase in sensitivity (e.g. {\it X-ray Surveyor} or similar future missions). 
In order to better constrain the Geminga PWN properties, particularly the stand-off distance ahead of the pulsar, observations of its forward bow shock would be very useful.  Bow shocks have been detected in H$_{\alpha}$ around nine pulsars \citep{Brownsberger2014}, but not around Geminga, presumably because the ISM is strongly ionized ahead of this 
pulsar. The recent first detection of a pulsar bow shock in far-ultraviolet around PSR J0437--4715 \citep{Rangelov2016} suggests that such shocks can be 
produced by supersonically moving pulsars even in the case of strong pre-shock 
ionization. Therefore, imaging of Geminga in the far-ultraviolet, which has not 
been done so far, could provide additional constraints on the PWN properties.

\acknowledgments
We thank Sandro Mereghetti and Eric Feigelson for helpful discussions regarding statistics of the axial-tail count distribution, and Konstantin Getman for discussion and valuable insights into the X-ray emission properties of late-type stars. We are also grateful to the anonymous referee for helpful suggestions.

Support for this work was provided by the National Aeronautics and Space Administration through Chandra Award Number G03-14057 issued by the Chandra X-ray Observatory Center, which is
operated by the Smithsonian Astrophysical Observatory for and on behalf of the National Aeronautics Space Administration under contract NAS8-03060.
This work was also partly supported by the National Aeronautics and Space
Administration under Grant Number NNX15AF10G issued through the Astrophysics Data Analysis Program.
A.M. Bykov acknowledges support from RSF grant 16-12-10225.

\bibliographystyle{aasjournal}
\bibliography{Gemingabib}

\appendix
\section{A. Assessment of the contamination in A4}
\label{starcontam}
There is a known optical/NIR source in segment A4 of the axial tail. PBZ10 reported the position and magnitudes from the USNO B1 catalog \citep{Monet2003}, the GSC2.3 catalog \citep{Lasker2008} and the 2MASS catalog \citep{Cutri2003} and found the colors to be consistent with a K star. They excluded an AGN as counterpart of the A4 emission based on the X-ray--to--optical flux ratio. 
Additional, new accurate magnitudes from the IPHAS DR2 catalog \citep{Barentsen2014}, $r=17.76 \pm 0.01$ and $i=17.19 \pm  0.01$, also constrain the star to be of a late K to early M spectral type using stellar colors of \citet{Covey2007}.
The object is classified as 'star' in the IPHAS DR2 catalog as well as in the UCAC4 Catalog \citep{Zacharias2013}.
\citet{Roeser2010} reported limits on the apparent proper motion of this object as $\mu_{\alpha} \cos{\delta}= -0.1 \pm 5.0$\,mas\,yr$^{-1}$ and $\mu_{\delta} = -3.4 \pm 5.0$\,mas\,yr$^{-1}$ with respect to observing epoch 1982.86.
Hence, the motion of this background star between epoch B and C1 (in both of which A4 is prominent) is negligible, and it could contribute to the X-ray flux of A4 in both epochs. There are, however, two arguments for a negligible contribution to A4's X-ray flux.\\

In epoch C1, the centroid position of the A4-emission (using a circle with $r=4\farcs{6}$) is $2.0\arcsec$ away from the IPHAS star position (epoch 2008), while a 2MASS star northeast of Geminga has an X-ray counterpart whose centroid position is only $0\farcs{35}$ away from its 2MASS position (the star also has a negligible proper motion). This northeast star is at a similar off-axis angle as A4, yet from its detected 16 counts we derived a $\sigma_{\rm centroid X}=0\farcs{49}$ and $\sigma_{\rm centroid Y}=1\farcs{1}$ (the star is very close to the gap between I3 and I2 chips), while the 33 counts of the A4-emission resulted in  $\sigma_{\rm centroid X}=1\farcs{58}$ and $\sigma_{\rm centroid Y}=1\farcs{71}$. The separation of the star from A4 and the achieved centroid position accuracy support the notion that the star is \emph{not} the counterpart of A4.
Furthermore, the A4 count distribution does not appear to be strongly centrally peaked in contrast to the one of the northeast star.\\ 
 
Another argument is based on the spectrum and temporal behavior of the A4 emission.
Since a typical star X-ray spectrum can be well described with, e.g., the APEC (Astrophysical Plasma Emission Code) model in Xspec, we can check whether a spectral fit of the A4 emission gives reasonable stellar parameters.
Including \emph{all} counts of A4 (note difference to Section~\ref{extem}), we obtained an APEC temperature of $kT=1.8\pm0.4$\,keV  in epoch C1 ($N_{\rm H}= 4 \times 10^{21}$\,cm$^{-2}$ set to Galactic $N_{\rm H \sc{I}}$ value \citep{Kalberla2005}; the temperature is higher at lower $N_{\rm H}$ values). 
This is a rather high temperature for a main-sequence K-to-M star, but it would be still consistent with emission from a young stellar object \citep{Getman2008,Guedel2004}. There is no prominent known star formation region within $1^{\circ}$ of Geminga, but we cannot entirely exclude a (diskless) young star. 
Since A4 is not prominent in epochs other than B and C, its emission would indicate an active state of the star -- if related to the star. Thus, the emission would be expected to be produced in flares. The expected duration of flares from an old star with a \emph{flare peak} temperature of $kT=1.8$\,keV would be between 1\,ks and 10\,ks (stellar flare compilation by \citet{Guedel2004}; see also Figure\,9 by \citealt{Getman2008}). For a young stellar object, the flare duration can be as long as 100\,ks (\citet{Getman2008}, their Figure\,9). 
Investigating  arrival times and energies of photons from A4 as well as from comparison test regions (background or other regions in the axial tail), we did not find any indication of flare behavior which could explain the A4 emission.
Instead, the A4 emission can be described as steady emission over the respective exposure times of epoch B (77\,ks) and C1 (62\,ks). This steady emission is similar to that seen in other (star-free) regions in the axial tail.\\

Based on the steady emission and -- with less emphasis (because of the low count statistics) -- on the spatial count distribution in A4, we conclude that the A4 emission is related to the PWN, with the star in A4 having a negligible effect on A4's X-ray emission properties.  
For a conservative spectral parameter estimate, we give in Table~\ref{extfit} also fit results for A4 where the star region was excluded. For the spatial analysis, however, we assume that the star has no significant influence on the count distribution of A4. 

\section{B. The geometrical shape spanned by the lateral tails}
\label{lattailsapp}
\begin{figure*}[b]
\noindent\begin{minipage}[b]{.5\textwidth}
\begin{center}
\includegraphics[width=75mm]{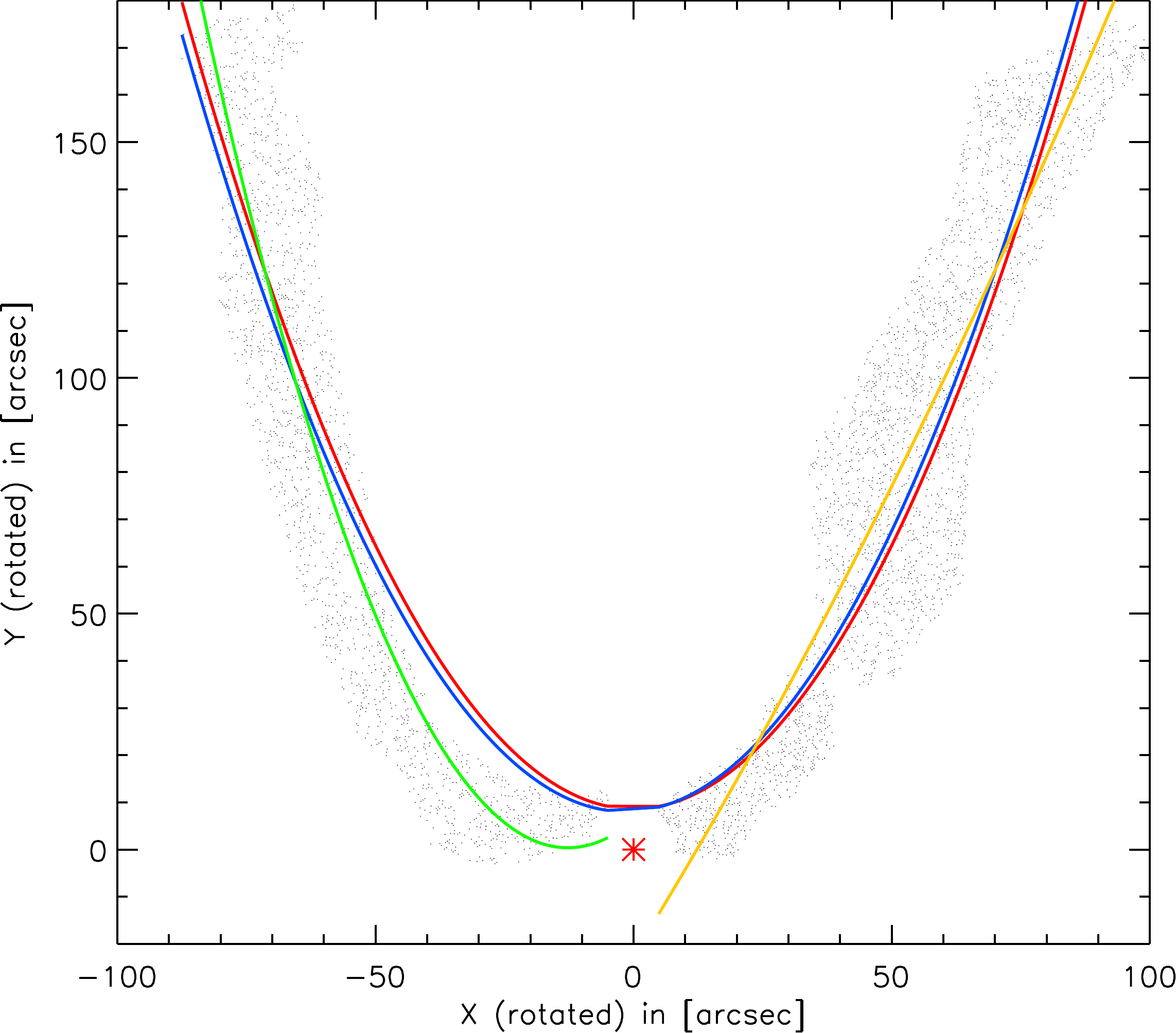}
\end{center}
\end{minipage} 
\begin{minipage}[b]{.5\textwidth}
\begin{center}
\includegraphics[width=75mm]{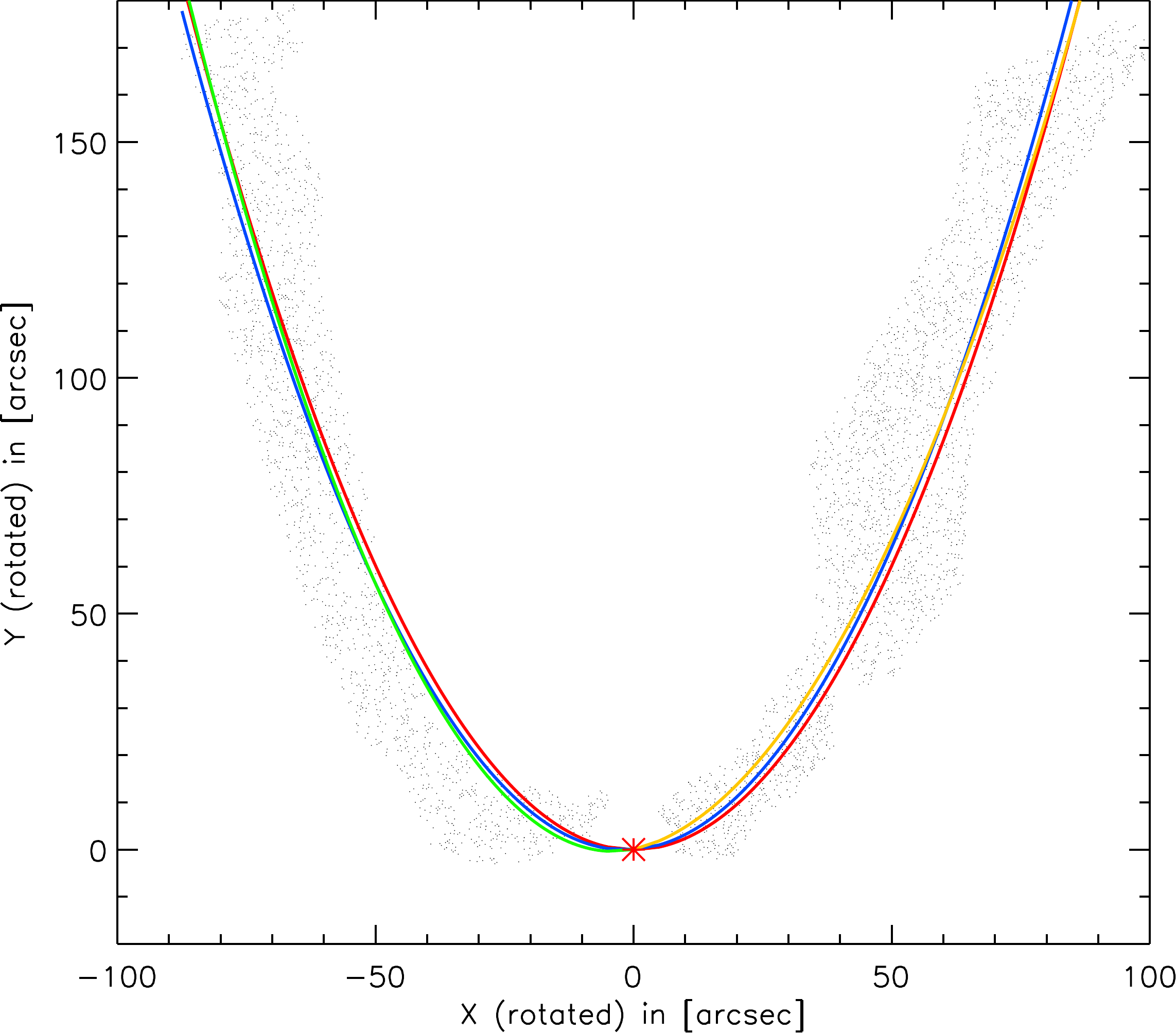}
\end{center}
\end{minipage}
\caption{The profile fits for the lateral tails. The small dots indicate the extracted events in the N- and S-tails regions (see Figure~\ref{Overview}). \protect\\
 \emph{Left panel:} The fits were done \emph{without} using the pulsar centroid position (marked as red dot). The red, blue, green, and yellow lines indicate the $y_0+a (x-x_0)^2$ fits with $x_0=0$ for the N\&S-tail, and with $x_0 \neq 0 $ for the N\&S-tail, the N-tail only, the S-tail only, respectively.\protect\\ 
\emph{Right panel:} These fits were done \emph{with} the pulsar centroid position (marked as red dot). The color choice is the same as in the left panel.}
\label{latfits}
\end{figure*}
We use a parabolic description for the shape spanned by the lateral tails.
It is a covenient shape to derive values for bending of potential jets or flux comparison. 
In order to obtain an approximate simple analytical expression for the geometry of the lateral tails, we extracted the regions of the N- and  S-tail from the merged 2012/2013 count image (the bin size is $1$\,ACIS pixel), and rotated and shifted it for the pulsar centroid to be at the zero point as seen in Figure~\ref{latfits}. 
Note that the region close to the pulsar is not included in order to avoid contributions of the axial tails and Bow. 
Pixels with $n$ number of counts ($n \leq 4$, in most cases $n=1$) are considered as $n$ individual points in our fitting procedure. Using the IDL routine \texttt{mpfitexpr} by Craig B. Markwardt, we fitted the count distribution of the lateral tails with the analytical expression  $y_0+a (x-x_0)^2$, both for the N\&S-tail together and for each tail individually. We do not attempt to estimate uncertainties of the parameters since the neglect of background counts and event localization errors in our method constitute a non-negligible oversimplification. 
Results are listed in Table~\ref{lattailfit} and shown in Figure~\ref{latfits}.
If we do not do not require the parabola to go through the pulsar position, the footpoint of the N\&S-tail fit is $8\farcs{6}$ away from the pulsar position. 
However, it is useful to also consider a fit with the pulsar centroid position.
This is due to the fact that we excluded the immediate surrounding of the pulsar because of potential contributions of the axial tails and Bow, but we observe the lateral tails connecting to the pulsar (Figures~\ref{tempchanges} and \ref{marxchanges}).
The N-tail and S-tail have slightly different shapes and require slightly different fits.
Though helical shapes could be potential underlying structures in the lateral tails we do not attempt to fit such curves due to low count numbers. 

\begin{deluxetable}{lrrrr}[h!]
\tablecaption{Fit results for the profile of the lateral tails\label{lattailfit}}
\tablewidth{0pt}
\tablehead{
\colhead{Parameter} & \colhead{$y_0+a x^2$} & \multicolumn{3}{c}{$y_0+a (x-x_0)^2$} \\
\colhead{} & \colhead{N\&S-tail} & \colhead{N\&S-tail} & \colhead{N-tail} & \colhead{S-tail}\\
\colhead{} & \colhead{red\tablenotemark{a}} & \colhead{blue\tablenotemark{a}} & \colhead{green\tablenotemark{a}} & \colhead{yellow\tablenotemark{a}}}
\startdata
\multicolumn{5}{c}{without pulsar centroid position}\\
\hline\\
$y_0$ & 8.622 & 8.037 & 0.372 & $-206.1$\\
$a$ & 0.022   & 0.022 & 0.036 & 0.004\\
$x_0$ & $-$   & $-1.6$ & $-12.8$ & $-207.1$\\
\hline\\
\multicolumn{5}{c}{with pulsar centroid position}\\
\hline\\
$y_0$ & $10^{-12}$ & $-0.062$ & $-0.393$ & $-0.857$\\
$a$ &  0.024 & 0.024 &  0.027 & 0.021\\
$x_0$ & $-$ & $-1.599$ & $-3.841$ & $-6.394$  \\
\enddata
\tablecomments{The units are arcsec for $x$ and $y$, thus arcsec for $y_0$ and $x_0$ and $1/{\rm arcsec}$ for $a$.}
\tablenotetext{a}{color in Fig.~\ref{latfits}}
\end{deluxetable}

\section{C. Implications for highly collimated jets}
\label{jetsapp}

\noindent 
Assuming a jet comprised of electrons with randomly oriented ultrarelativistic velocities and a magnetic field, the relativistic formula for the energy flow down a jet, $\dot{E_j}$, can be expressed \citep{Landau1959fluids} as 
\be
\label{equ1}
\dot{E_j}=\pi r^2_j v_j \Gamma^2 \left( \frac{4}{3} w_{\rm rel} + \frac{B^2}{4 \pi} \right),
\ee
where $\Gamma=(1-v^2_j / c^2)^{-1/2}$ is the bulk Lorentz factor,
$r_j=r_{j,16} 10^{16}$\,cm is the jet transverse radius,
$w_{\rm rel}$ is the energy density of the relativistic particles in the jet,
$B$ is the magnetic field.
In the following, $\rho$, $n$, and $v$ are the density, number density, and (bulk) velocity of either the jets (subscript $j$) or ambient medium (subscript $a$).\\ 

Using
the magnetization $k_m=w_{\rm mag}/w_{\rm rel}$ with $w_{\rm mag}=B^2 / (8\pi)$, and approximating for Geminga $B=20 k^{2/7}_m$\,$\mu$G (Formula~\ref{magformula} and Table~\ref{Bestimates}), we derive
\be
\dot{E_j}= 1.0\times 10^{32} r^2_{j,16} \frac{v_j}{c} \Gamma^2 (3 k^{4/7}_m + 2 k^{-3/7}_m) \text{erg s}^{-1}.   
\ee
The energy flow down a jet cannot be larger than the spin-down power of the pulsar, $\dot{E_j} =\xi_j \dot{E} < \dot{E}$. For a given $r_j$ and $v_j$, the energy flow is minimal at $k_m=0.5$.
We can therefore estimate an upper limit on the bulk flow velocity in the jet for Geminga, for $r_{j,16}=3.7$ ($10\arcsec$ at 250 pc), as $v_j < 0.9 c$. The velocity would be lower for more realistic $\xi_j$, e.g., $v_j < 0.42 c$ for $\xi_j=0.1$\\

The factor, $\Gamma^2 \left( \frac{4}{3} w_{\rm rel} + \frac{B^2}{4 \pi} \right)$ from equation~\ref{equ1} is the relativistic enthalpy per unit volume and its relation to the bending scale $R_b$ can be expressed as \citep{Odea1985}
\be
\frac{\Gamma^2 \left( \frac{4}{3} w_{\rm rel} + \frac{B^2}{4 \pi} \right) \frac{v^2_j}{c^2}}{R_b}= \frac{ \rho_a v^2_a}{r_j}.
\ee

Thus, for a bent jet (comprised of ultrarelativistic electrons) with the bending scale $R_b=R_{b,16} 10^{16}$\,cm, we further derive the ambient number density $n_a=\rho_a/\mu_H m_P$ ($m_P$ is the mass of a proton) as
\be
n_a = \frac{\xi_j \dot{E} }{\pi c \mu_H m_P R_b r_j v^2_a } \frac{v_j}{c}
=63 \frac{\xi_j \dot{E}_{35}}{R_{b,16} r_{j,16} v^2_{psr,7}} \frac{v_j}{c} \text{cm}^{-3}
\ee
Here, we identify $v_a$ with the pulsar space velocity $v_{psr}=v_{psr,7} 10^7$\,cm\,s$^{-1}$.
For Geminga's measured parameters, $\dot{E}_{35}=0.33$, $r_{j,16}=3.7$, $R_{b,16}=7.8$ ($1/(2a)=20.8\arcsec$ at 250 pc with $a=0.024$ from Appendix~\ref{lattailsapp}), $v_{psr,7}=2.11$, assuming $\mu_H \approx 1$, and applying the upper limit on $v_j/c$ from above, we obtain
$n_a=0.016 (\xi_j/0.1) \text{cm}^{-3} (v_j/c) < 0.015 (\xi_j/0.1) \text{cm}^{-3}$.
If we fix $\xi_j=0.1$, and thus $v_j < 0.42 c$, we obtain $n_a< 0.007 \text{cm}^{-3}$.
This is on the order of what one would expect for the number density of the hot ionized interstellar medium, potentially indicating the hot bubble blown by the Geminga supernova. Interestingly, the lack of any H$_\alpha$ emission from the forward shock also implies a highly ionized medium.
A denser medium would bend the jets stronger than what is observed.
We note that in such a case of low ambient medium density one could expect to see any equatorial outflow of the pulsar well ahead of it. However, no prominent emission is oberserved there.
Further investigation, e.g., an MHD model of the shocked flow and measured ISM densities at the position of Geminga are needed to further constrain physical properties of the here outlined highly collimated jet interpretation.
  
\end{document}